\documentclass[8pt]{article}
\usepackage{graphicx}
\usepackage{pgfplots}
\usepackage{newlfont}
\usepackage[colorlinks,linkcolor=blue,citecolor=blue,anchorcolor=blue,bookmarksopen,pdfpagetransition={Wipe}]{hyperref}
\usepackage{amsmath,amsthm,amssymb}
\usepackage{subfigure}
\usepackage{amsmath}
\usepackage[linesnumbered,ruled]{algorithm2e}
\newtheorem{thm}{Theorem}[section]

\numberwithin{equation}{section}
\begin{document}
\title{\textbf{A combination chaotic system and application in color image encryption }}
\author{R. Parvaz$^a$\footnote{Corresponding author:  rparvaz@uma.ac.ir }\
, M. Zarebnia$^a$\footnote{zarebnia@uma.ac.ir}}
\date{}
\maketitle
\begin{center}
$^a$Department of Mathematics, University of Mohaghegh Ardabili,
56199-11367 Ardabil, Iran.\\
\end{center}
\begin{abstract}
In this paper, by using Logistic, Sine and Tent systems we define a combination chaotic system.
Some properties of the chaotic system are studied by using figures and numerical results.
A color image encryption algorithm is introduced based on new chaotic system.
Also this encryption algorithm can be used for gray scale or binary images.
 The experimental results
of the encryption algorithm  show that the encryption algorithm
is secure and practical.
\indent
\end{abstract}
\vskip .3cm

\section{Introduction}
\indent \hskip .65cm

Image plays an important role in the data transfer. With rapid development
network communication, image security has become
increasingly important. The first step in chaotic encryption was introduced by Matters \cite{1}.
In recent years,
much attention has been given in the literature to the
development, analysis and implementation of chaotic system for the image and the data encryption;
see, for example, \cite{2,3,4,5,6}.
Chaos maps as Logistic map, Sine map and Tent map are used in image encryption
algorithm because chaotic
maps have high sensitivity to their initial values
and control parameters  \cite{7,8}.
Logistic,  Sine  and Tent maps have some disadvantages. These maps
for some values of $r$ have chaotic behavior. Also these maps have non-uniform distribution
over output. Many methods have been proposed to solve these problems, for example see  \cite{9}.
To overcome these problems, in this paper by using different functions as $\sin(x), \cos(x), \ldots$, we combine
Logistic (or Sine) map and Tent map  then by using this combination the different
chaotic systems can be found. In the next step by using combination map, {\scriptsize XOR} operation and  circ shift the
color image encryption algorithm is introduced. In this encryption algorithm
in the first step, color image is divided into twelve parts and then by using
combination map the encryption process for each of the parts is done, in the last step,
we combine parts and then the encryption process combination image is repeated. \\

The organization of this paper is as follows: In Section 2, Logistic-Tent
combination map is explained. In Section 3, we present
the color image encryption algorithm. Simulation results and security analysis
are given in Section 4. A
summary is given at the end of the paper in Section 5.

\section{A combination chaotic system}
\indent \hskip .65cm
In this section, we describe our combination of chaotic systems.
Logistic,  Sine and Tent maps are defined as follows
\begin{align}
&x_{n+1}=L(r,x_n):=rx_n(1-x_n),\\
&x_{n+1}=S(r,x_n):=r\sin(\pi x_n)/4,
\end{align}
\begin{align}
x_{n+1}=T(r,x_n):=
\left\{%
\begin{array}{ll}
rx_n/2,&when~x_n<0.5,\\
\\
r(1-x_n)/2,&when~x_n\geq0.5.\\
\end{array}%
\right.
\end{align}

where parameter $r\in(0,4]$. It is known that the Logistic system
(Sine or Tent system) for some values of $r\in(0,4]$ has chaotic behavior.
Fig \ref{f1-a} and Fig \ref{f3-a}  show that the Logistic system for $r=2$ has not chaotic behavior.
To overcome the above problem, the combination of chaotic system as Logistic Tent system (LTS) introduced in \cite{9}.
Histogram of the Logistic Tent system is showed in Fig \ref{f2-a}.
Chaotic range is not limited for  the Logistic Tent system but
from Fig \ref{f2-a} we can see that the histogram of the Logistic Tent system is not flat enough. Non-uniform distribution over output series leads to weakness in the statistical attack. For solving this problem we add weights and functions in the Logistic Tent system or
the Sine Tent system
as follows
\begin{align}\label{e1}
&x_{n+1}=G_r(x_n):=\nonumber\\
&\left\{%
\begin{array}{ll}
\omega_1f_1 \circ F(r,x_n)+\alpha_1g_1(rx_n)+\xi_1\frac{(\beta_1-r)x_n}{2}~mod~1,&when~x_n<0.5,\\
\\
\omega_2f_2 \circ F(r,x_n)+\alpha_2g_2(rx_n)+\xi_2\frac{(\beta_2-r)(1-x_n)}{2}~mod~1,&when~x_n\geq0.5,\\
\end{array}%
\right.
\end{align}
 where $F(r,x_n)$ is Logistic or Sine map. In (\ref{e1}), $f_i(x)$ and $g_i(x)~(i=1,2)$ can be considered as $ax, \sin(ax),$ $\cos(ax),\tan(ax),$ $\cot(ax), \exp(ax), \log(ax)$ (where $a$ is real constant) and any other appropriate function.
Also in the above formula, $\omega_i,$ $ \alpha_i, \xi_i$ and $\beta_i~(i=1,2)$ are real numbers and parameter
 $r\in(0,4]$. In next step to investigate the properties of the new system, we consider the following cases
\begin{itemize}
\item i: $2\omega_1=\omega_2=20,-\alpha_1=\alpha_2=-2, 2\xi_1=\xi_2=4, \beta_1=4,$ $\beta_2=-20,
f_1(x)=\sin(x), f_2(x)=\exp(x),g_1(t)=\cot(x), g_2=\cos(\pi x), F(r,x_n)=L(r,x_n)$.
\\
\item ii: $\omega_1=\omega_2=20,-\alpha_1=\alpha_2=0, 2\xi_1=\xi_2=1, \beta_1=\beta_2=80,$ $
f_1(x)=\sin(x), f_2(x)=\exp(x), F(r,x_n)=L(r,x_n)$.
\\
\item iii: $\omega_1=\omega_2=1, \alpha_1=\alpha_2=1,  \xi_1=7, \xi_2=15, \beta_1=2\beta_2=40,$ $
f_1(x)=\cos(x), f_2(x)=\tan(x),g_1(x)=\tan(x), g_2(x)=x, F(r,x_n)=S(r,x_n)$.
\end{itemize}
For discrete time system as (\ref{e1}), the Lyapunov exponent for an orbit starting with $x_0$ is defined as follows
\begin{equation}
LE(x_0,r):=\lim_{n\rightarrow\infty} \frac{1}{n}\sum^{n-1}_{i=0}\ln|G_{r}^{\prime}(x_i)|.
\end{equation}
The degree of "sensitivity to initial conditions" can be measured by using the Lyapunov exponent.
In \cite{10} for the Lyapunov exponent the following theorem is mentioned.
\begin{thm}\label{th1}
If at least one of the average Lyapunov exponents is positive, then the
system is chaotic; if the average Lyapunov exponent is negative, then the orbit is
periodic and when the average Lyapunov exponent is zero, a bifurcation occurs.
\end{thm}
\begin{center}
\begin{figure}
\centering
\includegraphics[width=1.1\textwidth]{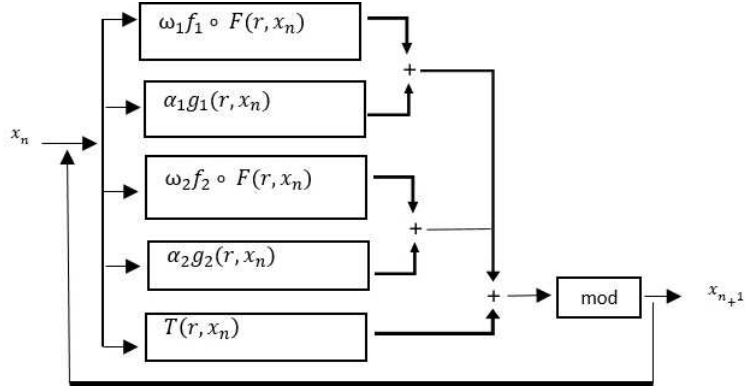}
\caption{{\footnotesize
The combination chaotic system.
} }
\label{f10}
\end{figure}
\end{center}

The Lyapunov exponent plot for Case (i) is given in Fig \ref{f1-b}. From this
figure we can see that for some values of $r$ the Lyapunov exponent are negative.
The Bifurcation diagram for the Case (i) has been shown in Fig. \ref{f4-b}. From this figure
we can see that white lines appear in places where
the Lyapunov exponent are negative.
Also for Case (ii)  and Case (iii) from Fig.s \ref{f1-c}-\ref{f1-d} we can say that for
all values of $r\in(0,4]$
the Lyapunov exponent are positive. Also the Cobweb plots \ref{f3-b}, \ref{f3-c} and \ref{f3-d}
show chaotic behavior for the Case (i), Case (ii) and the Case (iii).
The Fig.s \ref{f2-c}-\ref{f2-d} show that the Case (ii) and Case (iii) have uniform
distribution over  output range. Also from Fig. \ref{f3-b} we can see that
the Case (i) has not uniform distribution over its output range.
Two orbits of the Case (i) and Case(ii) are shown in Figs. \ref{f4-a}-\ref{f4-b}. By using this figure we can see that
the Case (i) and Case(ii) are  much more sensitive to the starting points. By using
the Lyapunov Exponents figures we can say that
the Lyapunov Exponents of two
examples are all larger than the Logistic map. Also the distribution of new combination chaotic system is more
uniform than the distribution Logistic Tent system.
Then by using suitable functions and parameters,  appropriate chaotic system can be found.\\
In the next section, as application  of the proposed chaotic
system, we  introduce an
image encryption algorithm.
\begin{figure}[!ht]\label{f1}
\centering
\subfigure[{}]{\label{f1-a}
\includegraphics*[width=.35\textwidth]{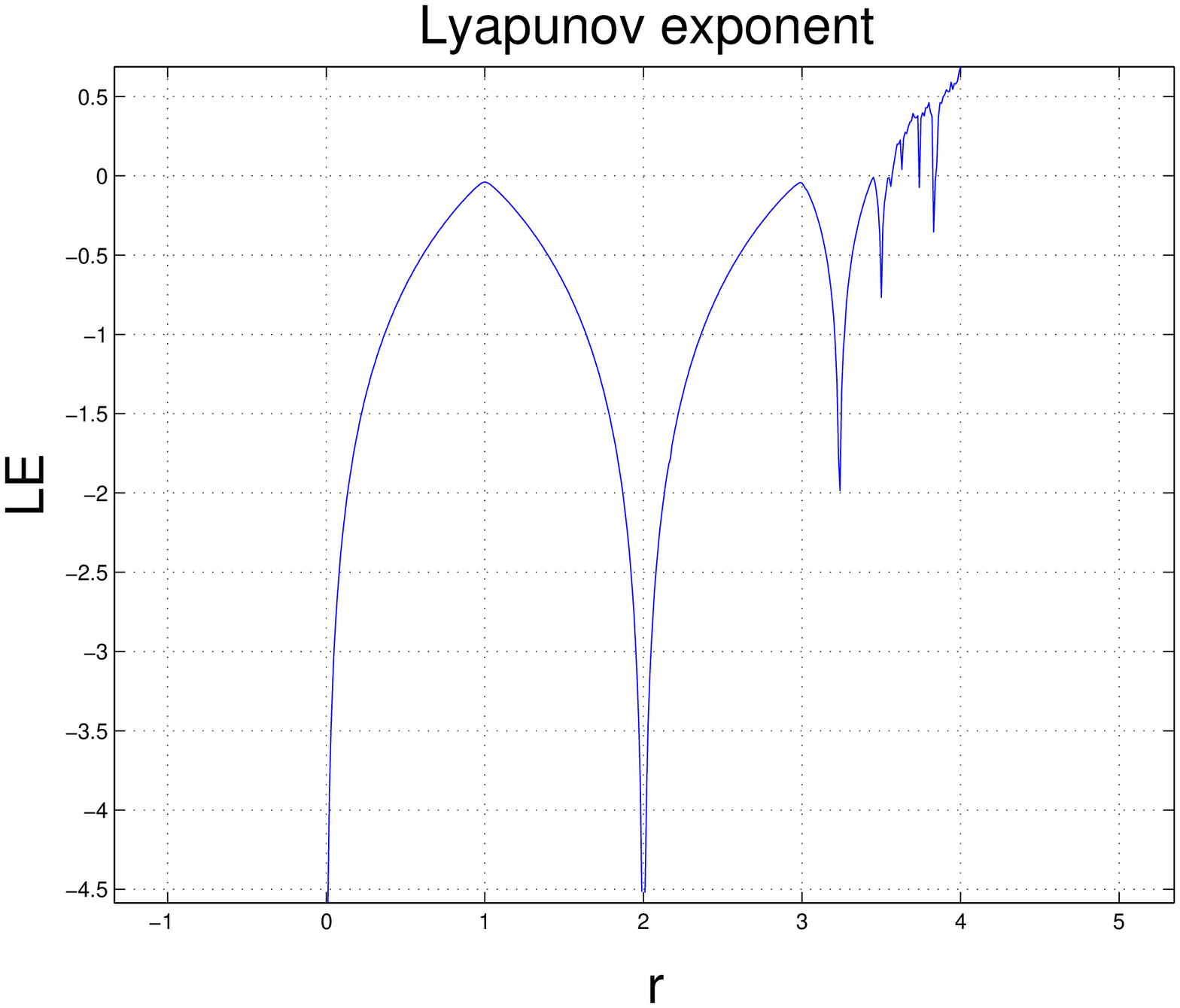}}
\hspace{2mm}
\subfigure[{}]{\label{f1-b}
\includegraphics*[width=.35\textwidth]{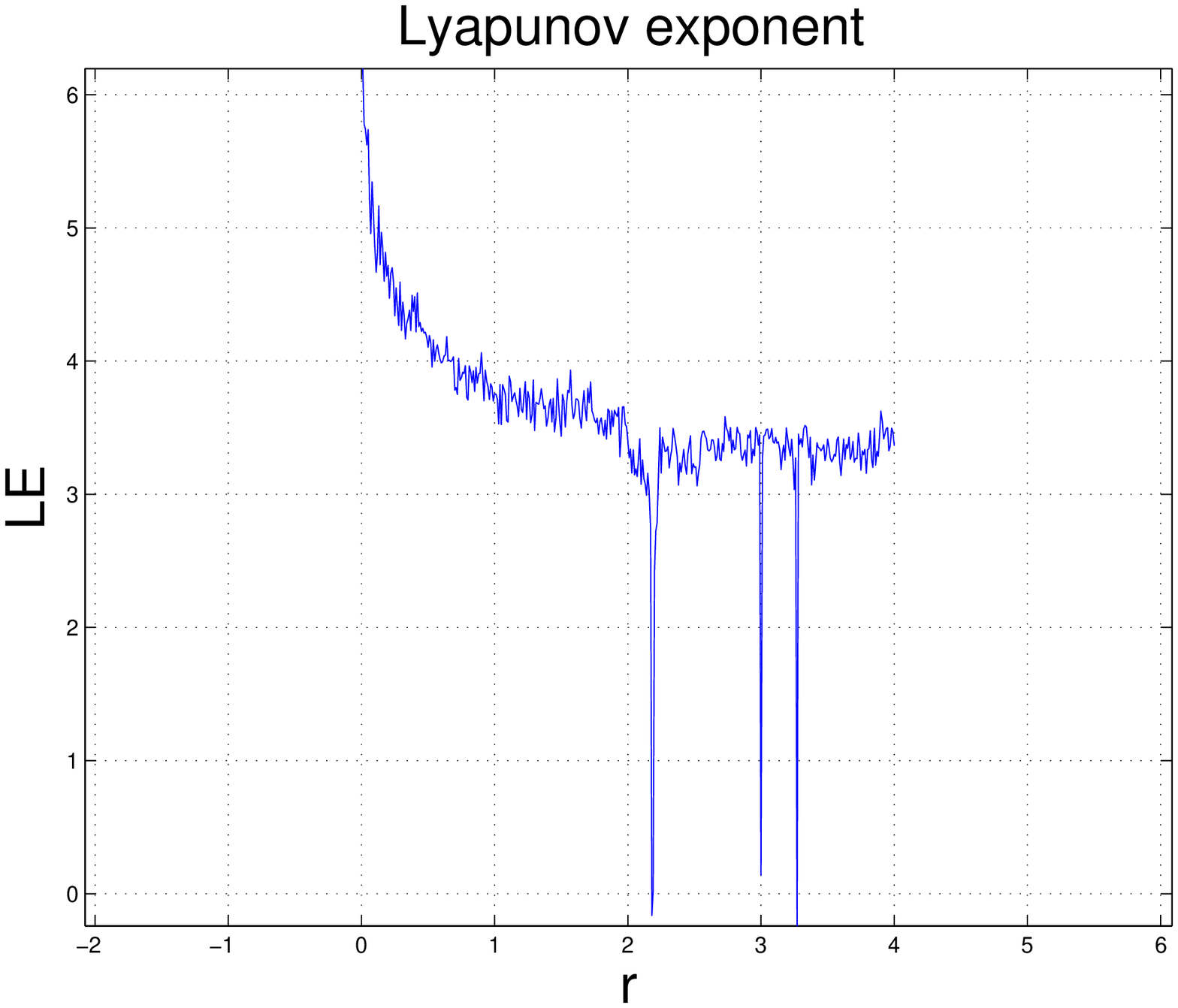}}
\subfigure[{}]{\label{f1-c}
\hspace{2mm}
\includegraphics*[width=.35\textwidth]{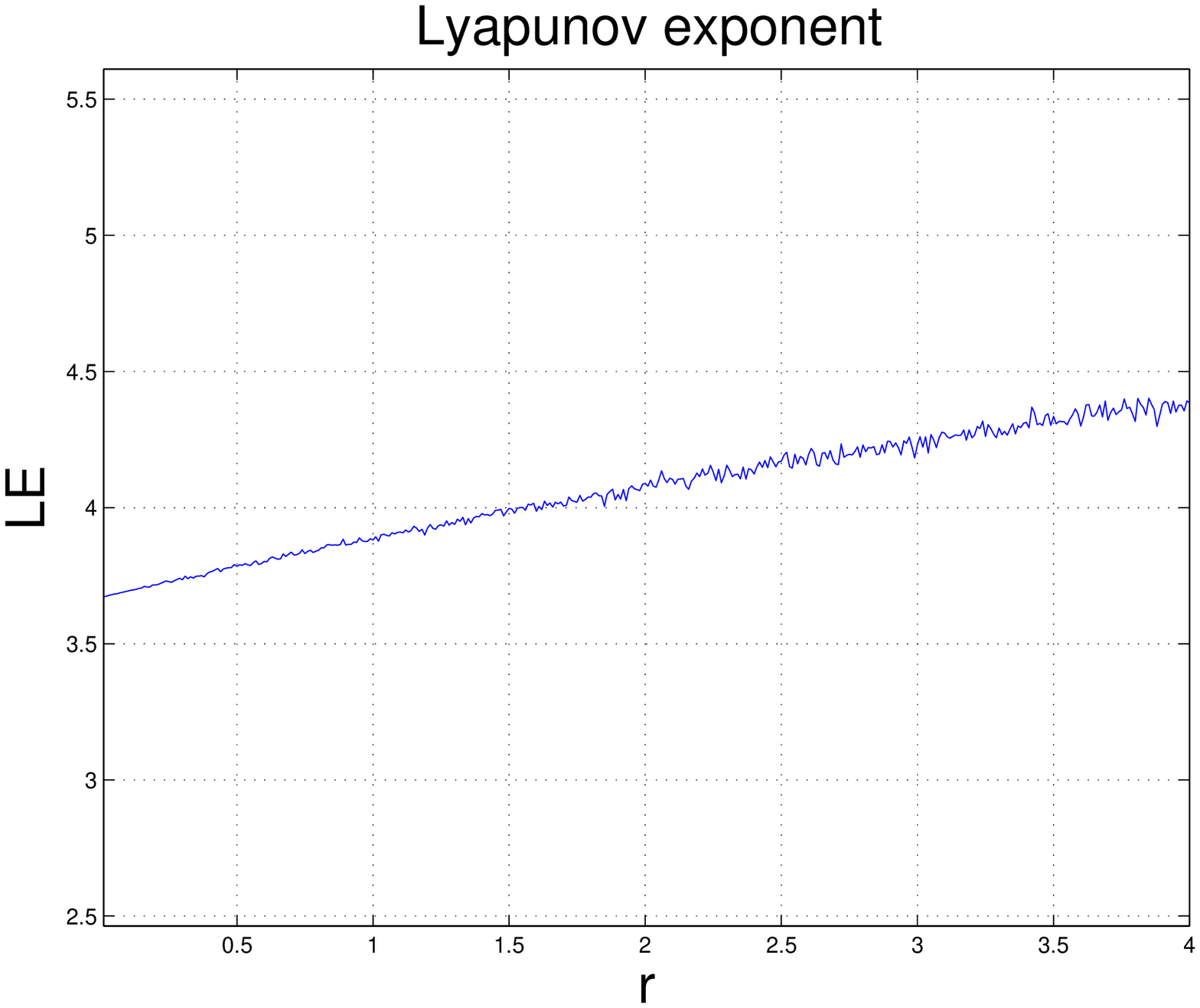}}
\hspace{2mm}
\subfigure[{}]{\label{f1-d}
\includegraphics*[width=.35\textwidth]{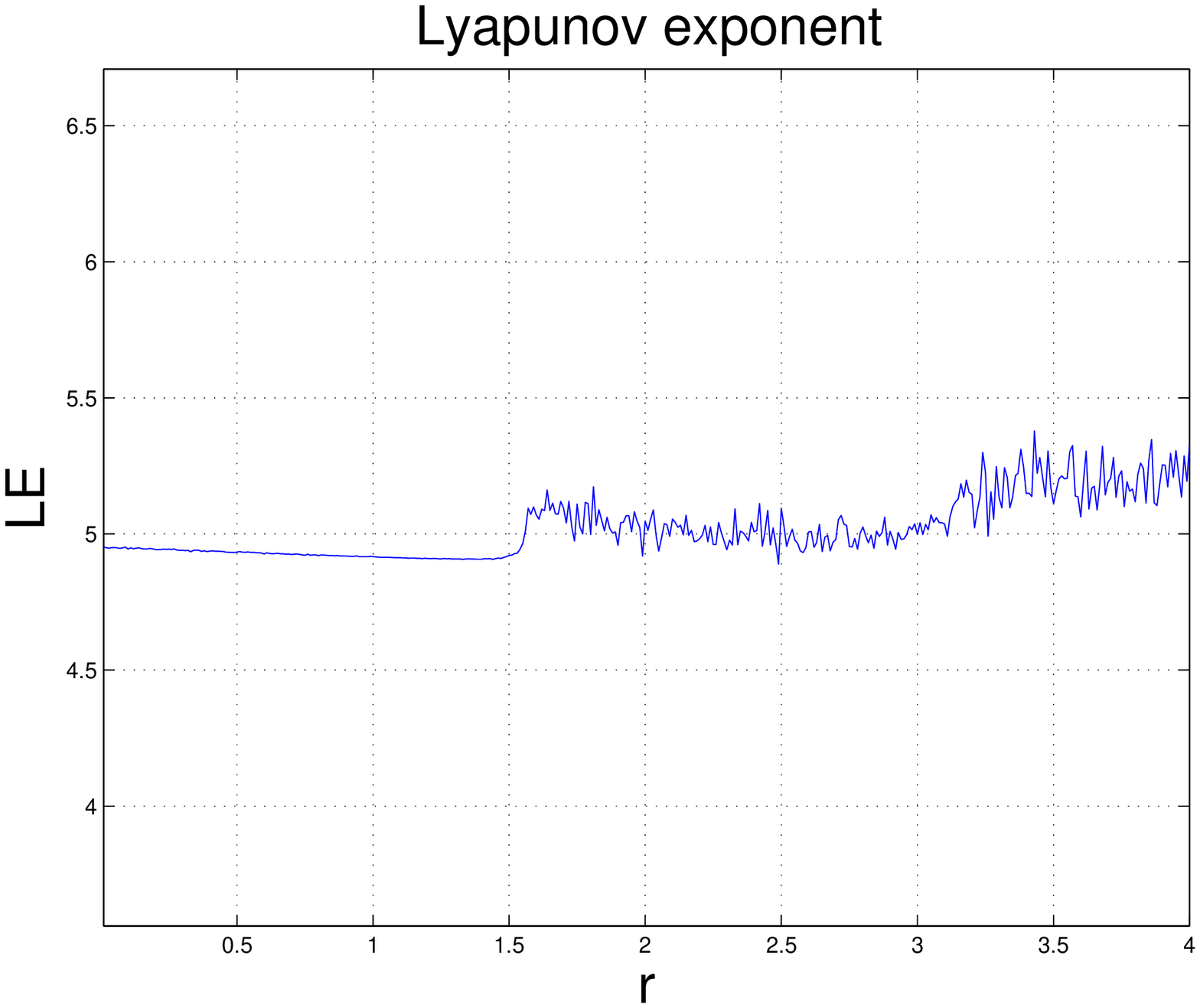}}
\caption{{\footnotesize
 Lyapunov exponent plot for the (a) Logistic map, (b) Case (i),
(c) Case (ii), (d) Case (iii).
}}
\end{figure}

\begin{figure}[!ht]\label{f2}
\centering
\subfigure[{}]{\label{f2-a}
\includegraphics*[width=.35\textwidth]{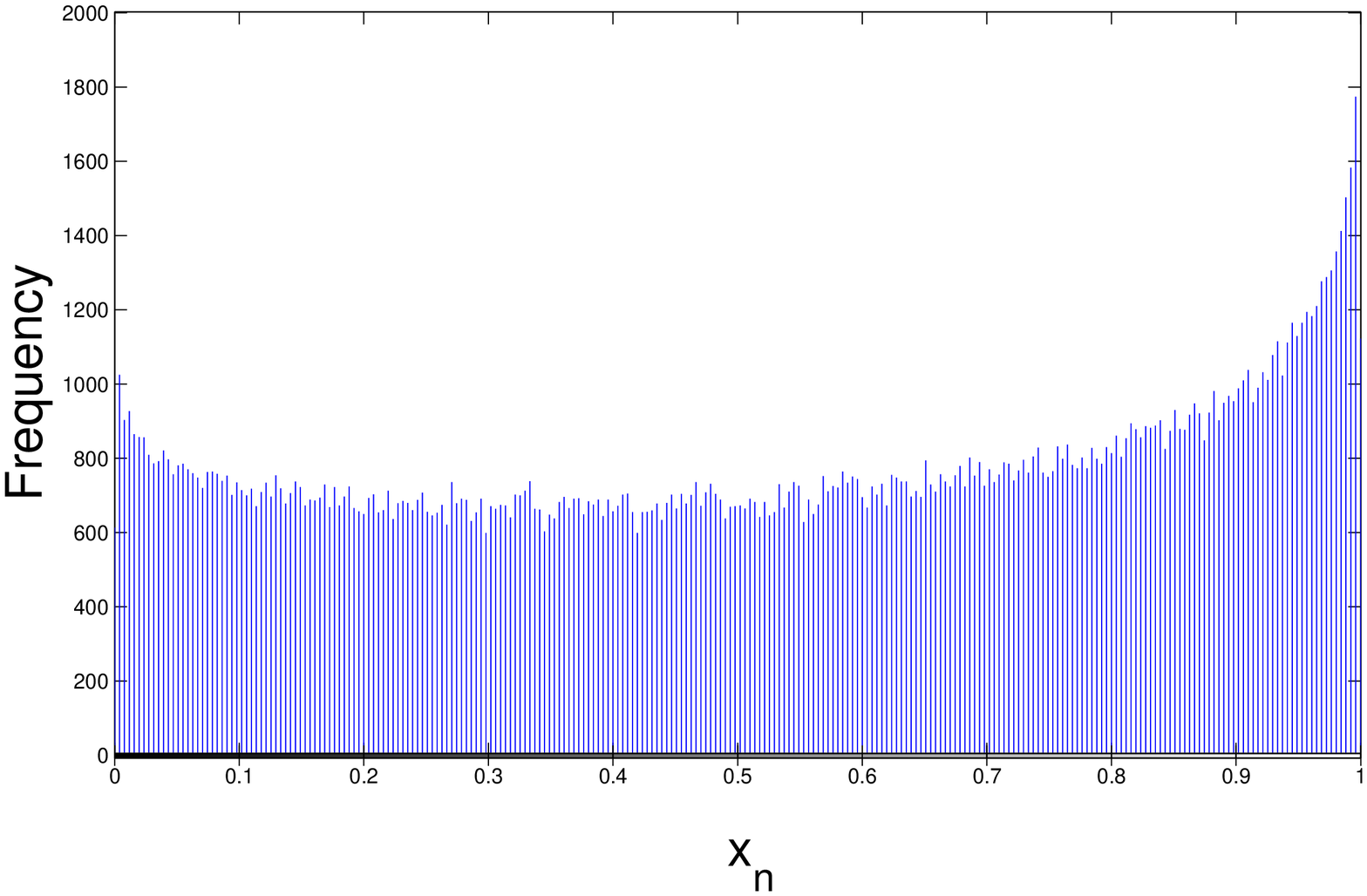}}
\hspace{2mm}
\subfigure[{}]{\label{f2-b}
\includegraphics*[width=.35\textwidth]{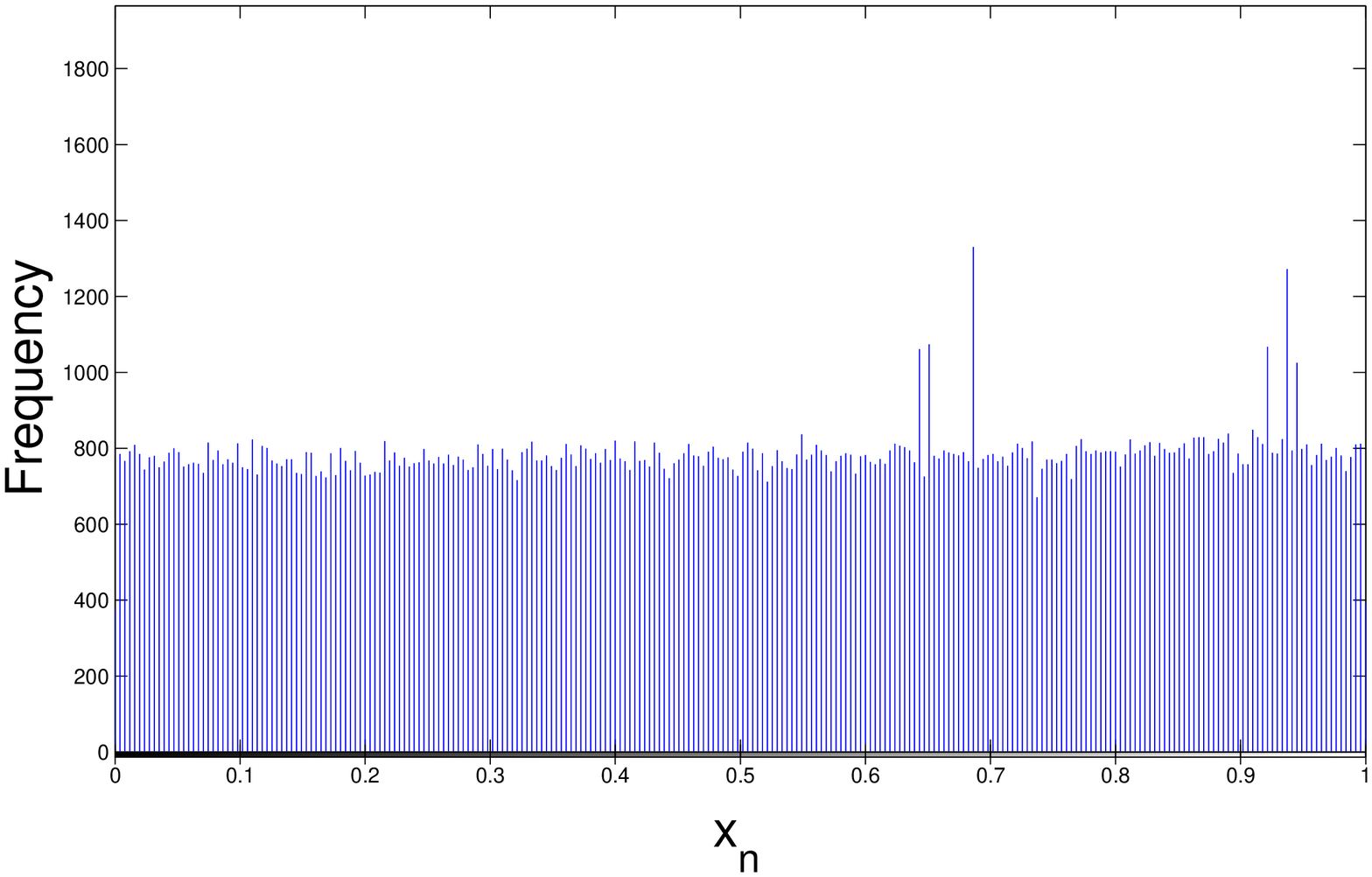}}
\subfigure[{}]{\label{f2-c}
\hspace{2mm}
\includegraphics*[width=.35\textwidth]{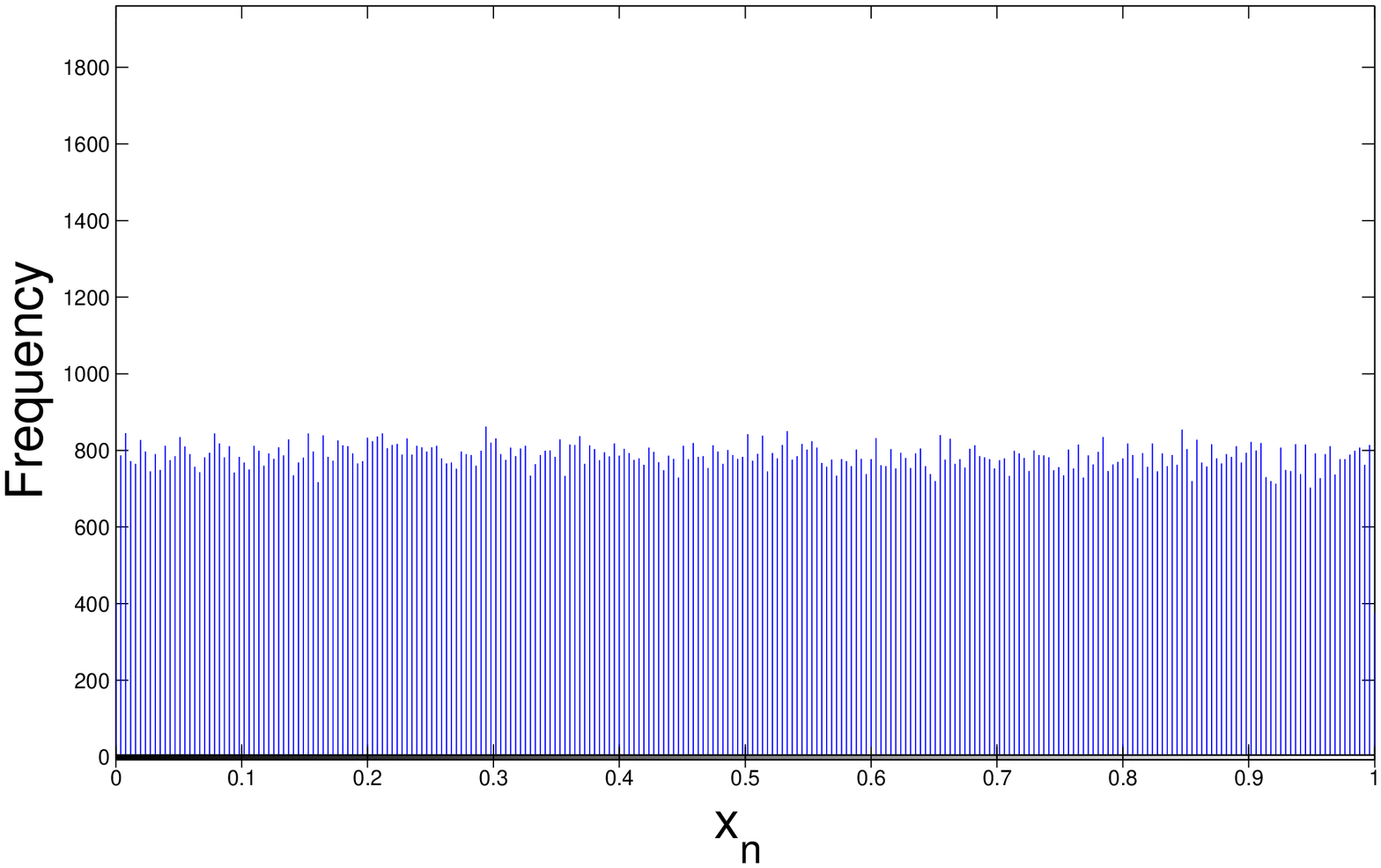}}
\hspace{2mm}
\subfigure[{}]{\label{f2-d}
\includegraphics*[width=.35\textwidth]{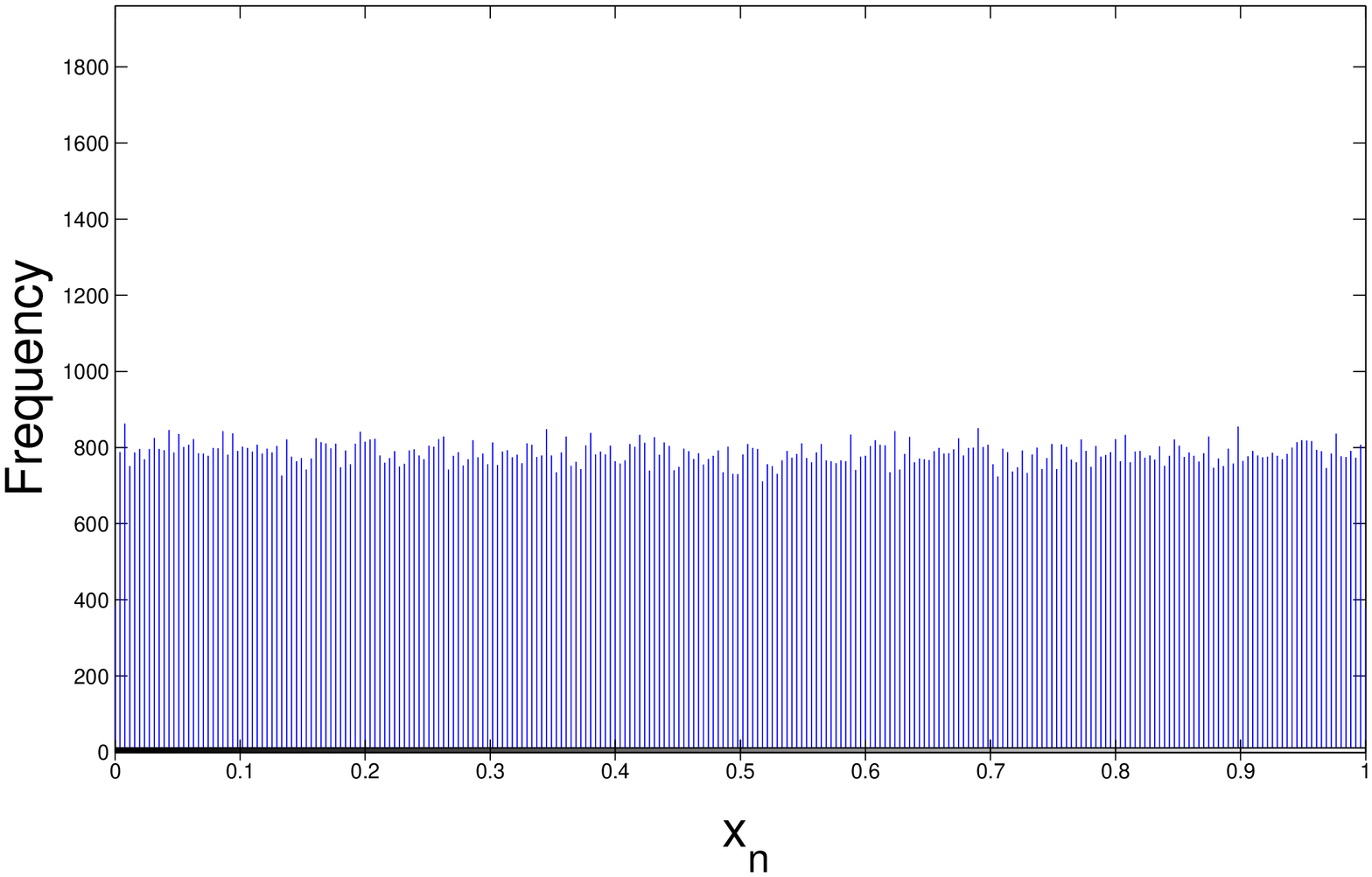}}
\caption{{\footnotesize
Histogram of the (a) Logistic Tent map, (b) Case (i),
(c) Case (ii), (d) Case(iii).
}}
\end{figure}

\begin{figure}[!ht]\label{f3}
\centering
\subfigure[{}]{\label{f3-a}
\includegraphics*[width=.35\textwidth]{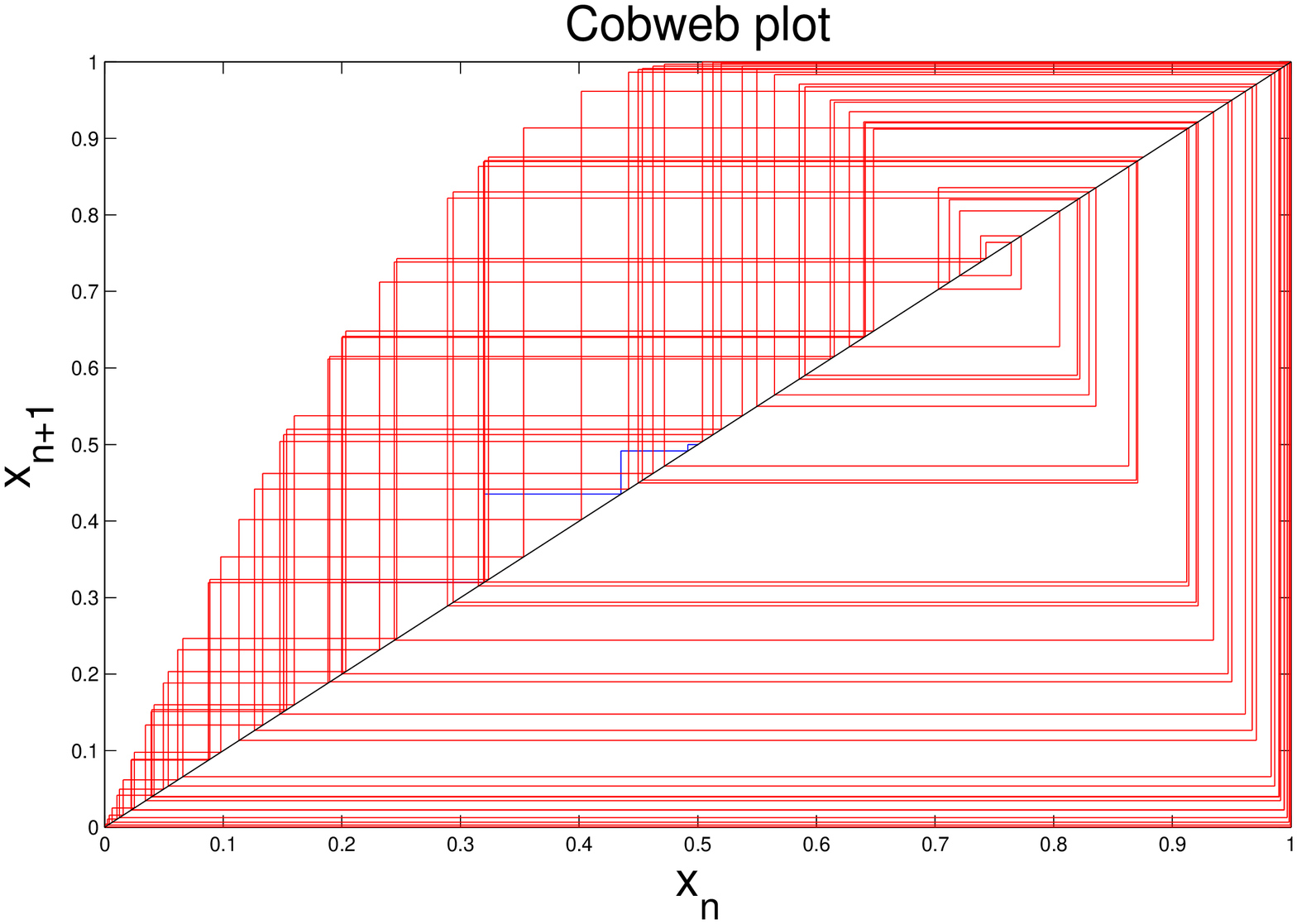}}
\hspace{2mm}
\subfigure[{}]{\label{f3-b}
\includegraphics*[width=.35\textwidth]{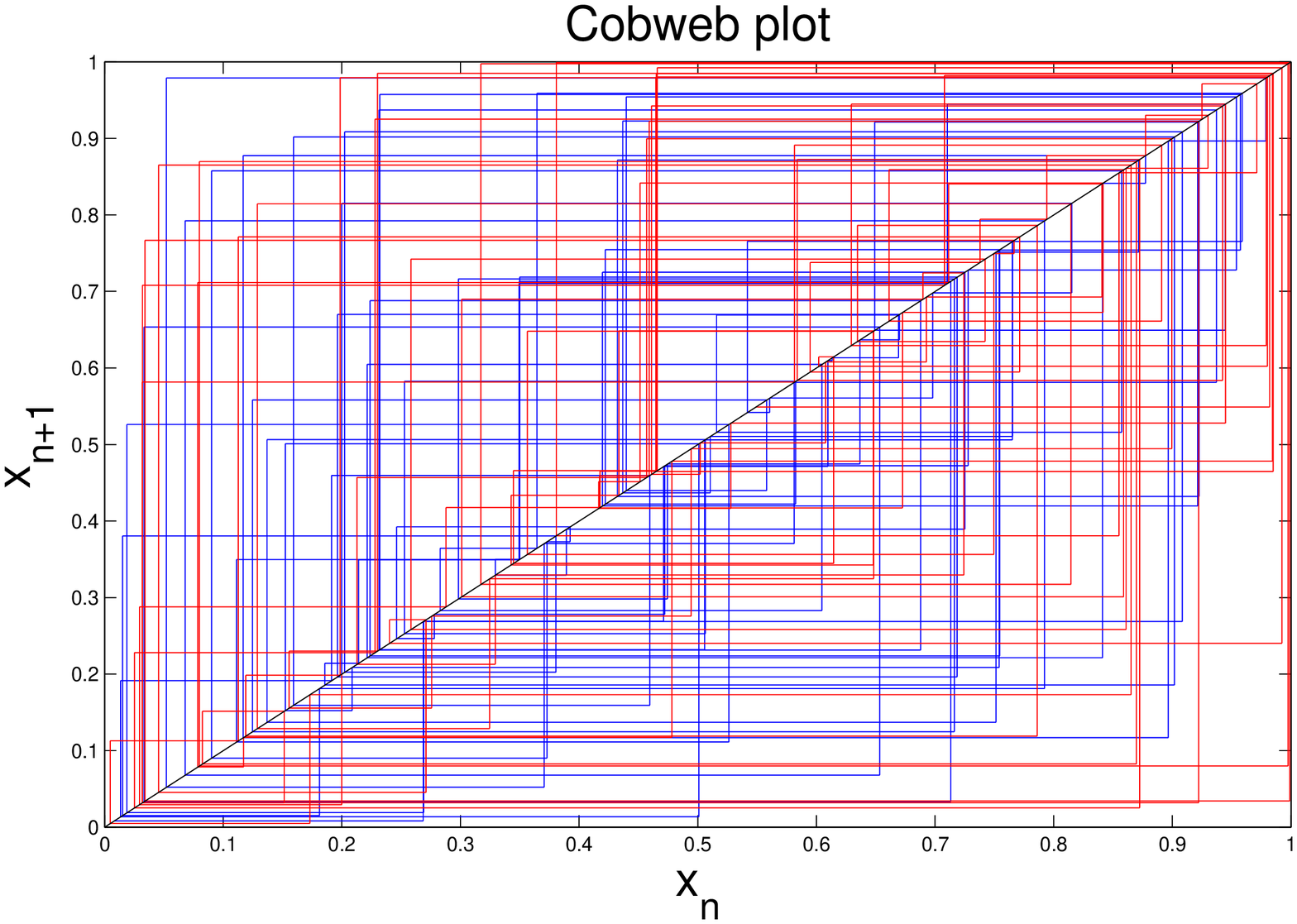}}
\subfigure[{}]{\label{f3-c}
\hspace{2mm}
\includegraphics*[width=.35\textwidth]{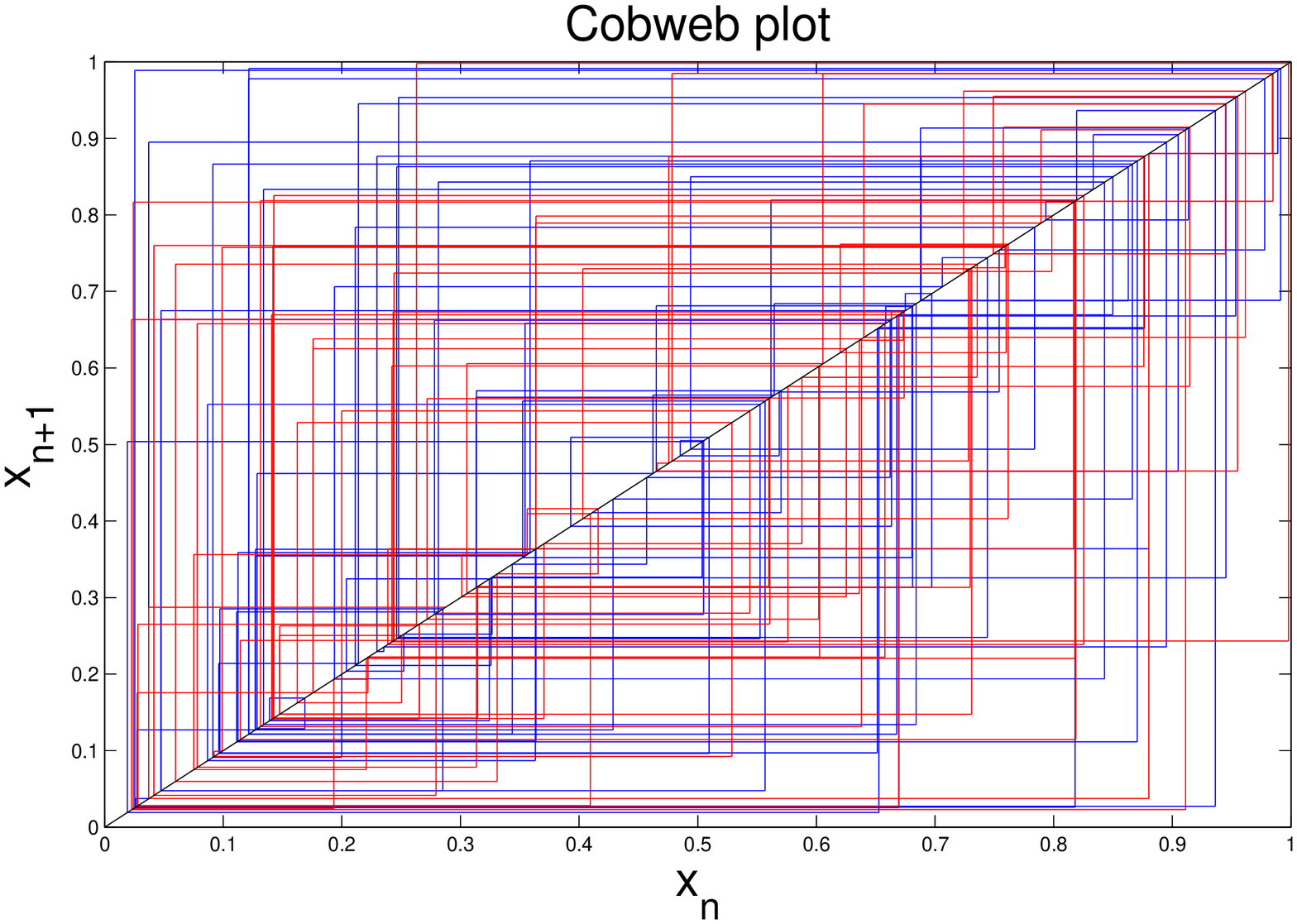}}
\hspace{2mm}
\subfigure[{}]{\label{f3-d}
\includegraphics*[width=.35\textwidth]{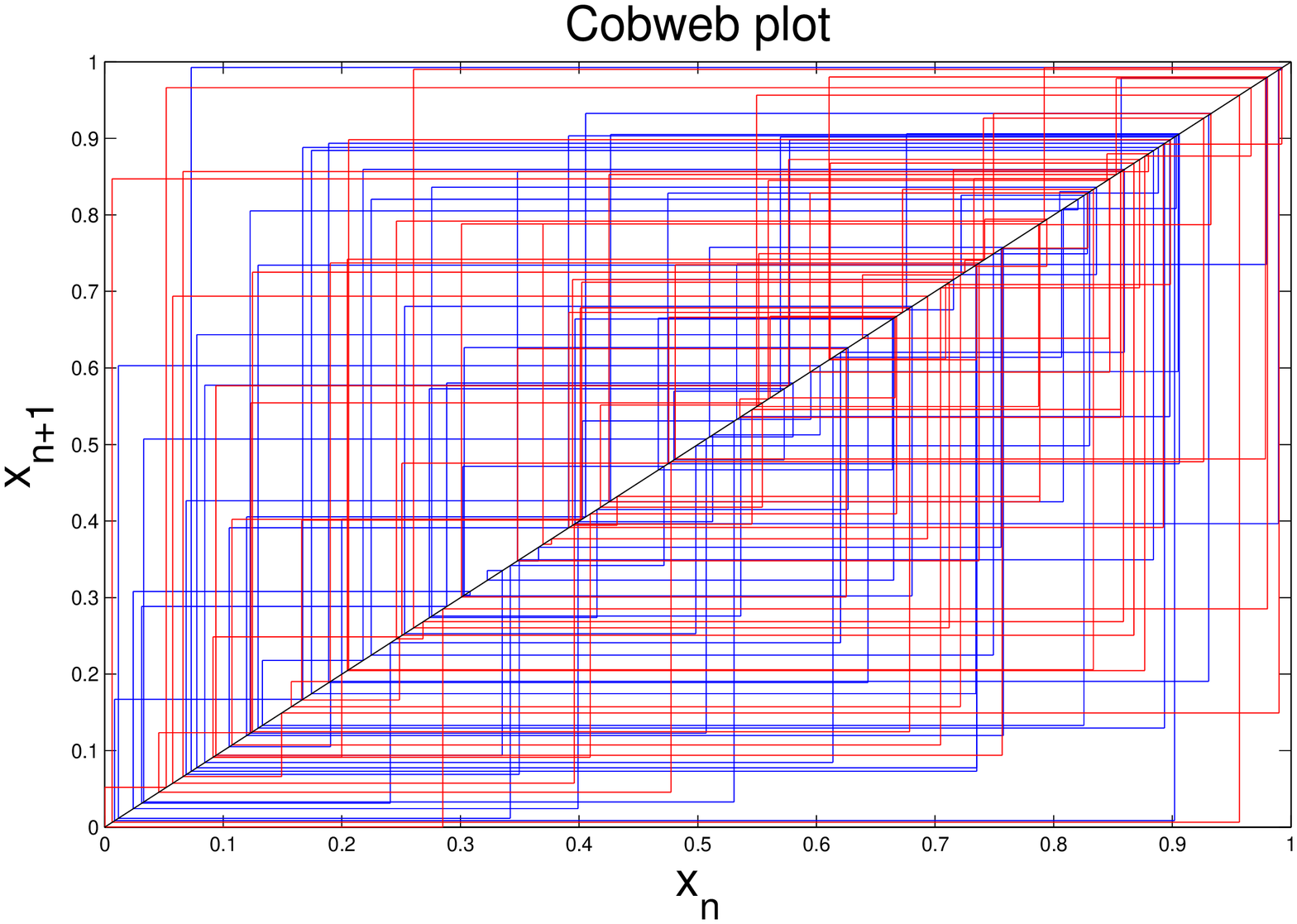}}
\caption{{\footnotesize
Cobweb plot for the  (a)  Logistic map, (b) Case (i),
(c) Case (ii), (d) Case(iii) (blue: r=2, red: r=4).
}}
\end{figure}
\begin{figure}[!ht]\label{f4}
\centering
\subfigure[{}]{\label{f4-a}
\includegraphics*[width=.35\textwidth]{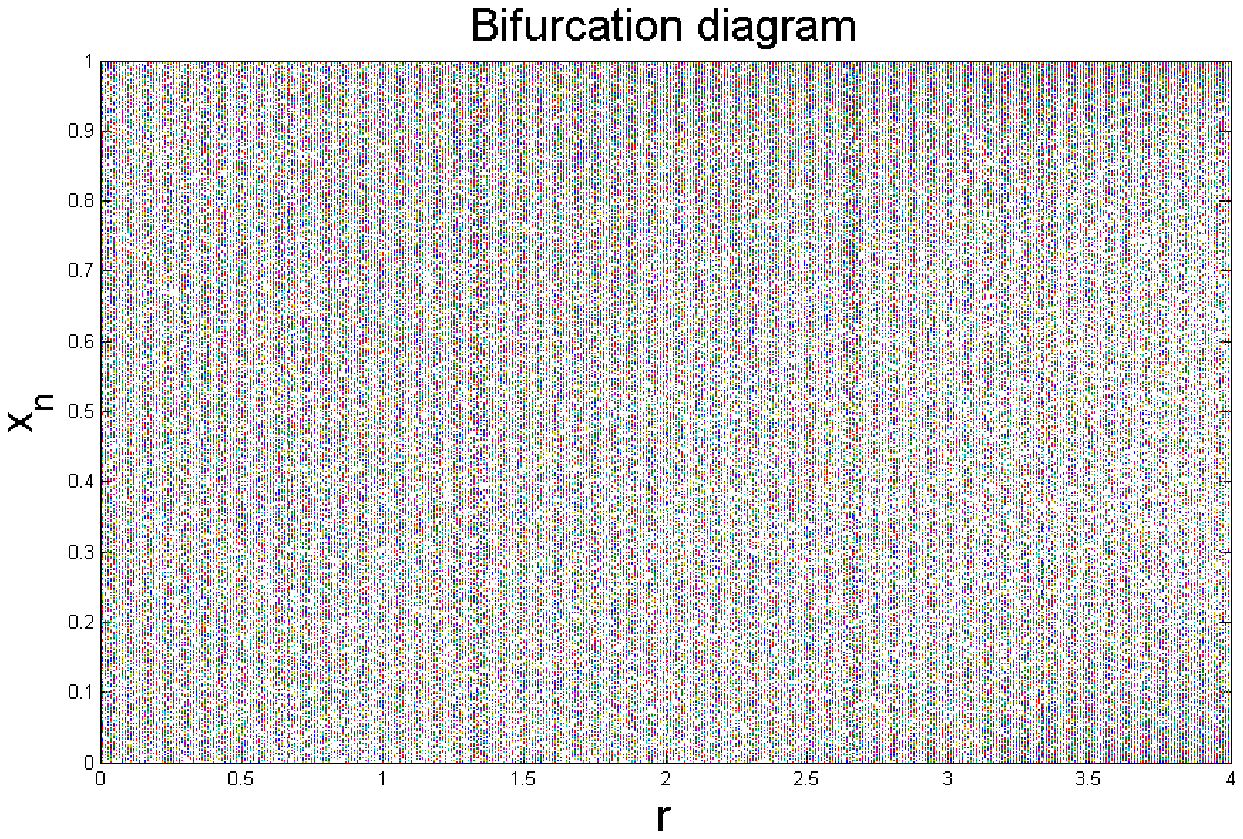}}
\hspace{2mm}
\subfigure[{}]{\label{f4-b}
\includegraphics*[width=.35\textwidth]{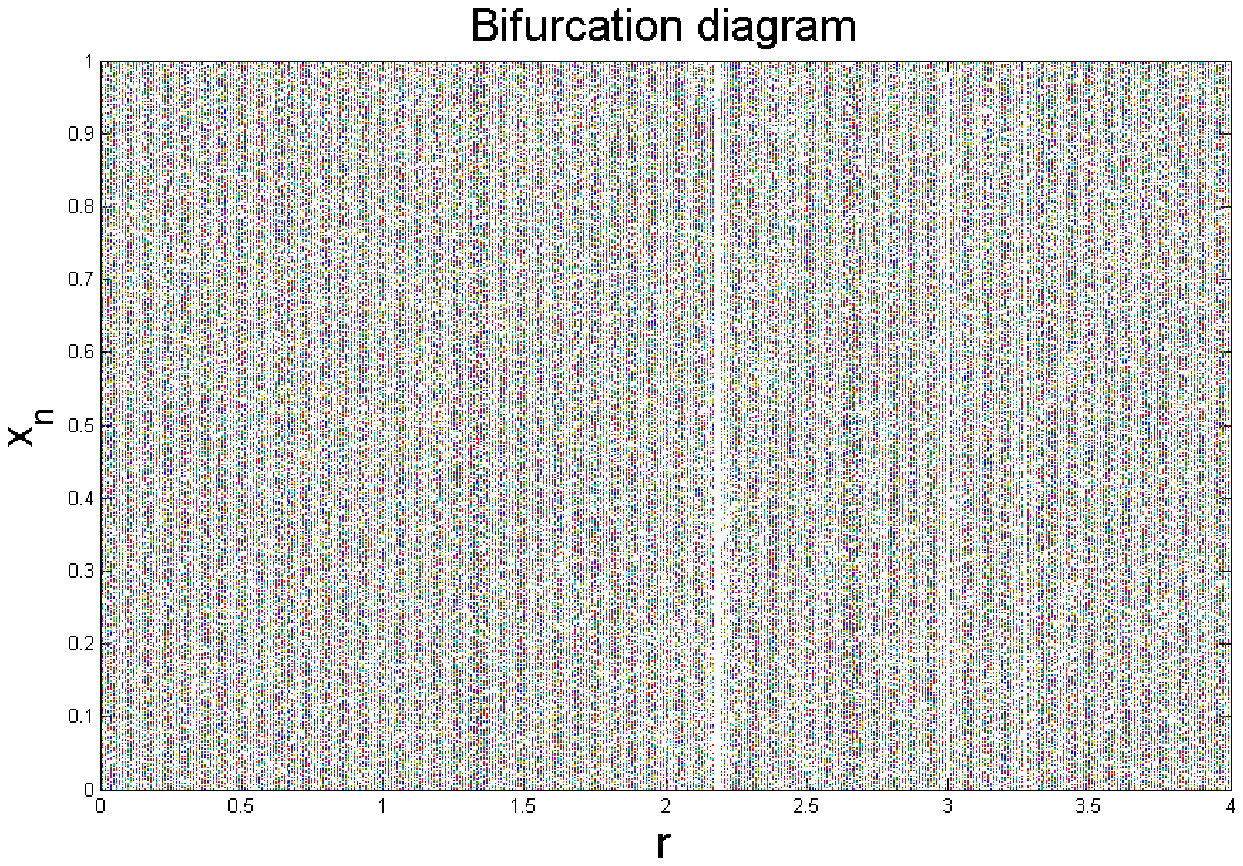}}
\subfigure[{}]{\label{f4-c}
\hspace{2mm}
\includegraphics*[width=.35\textwidth]{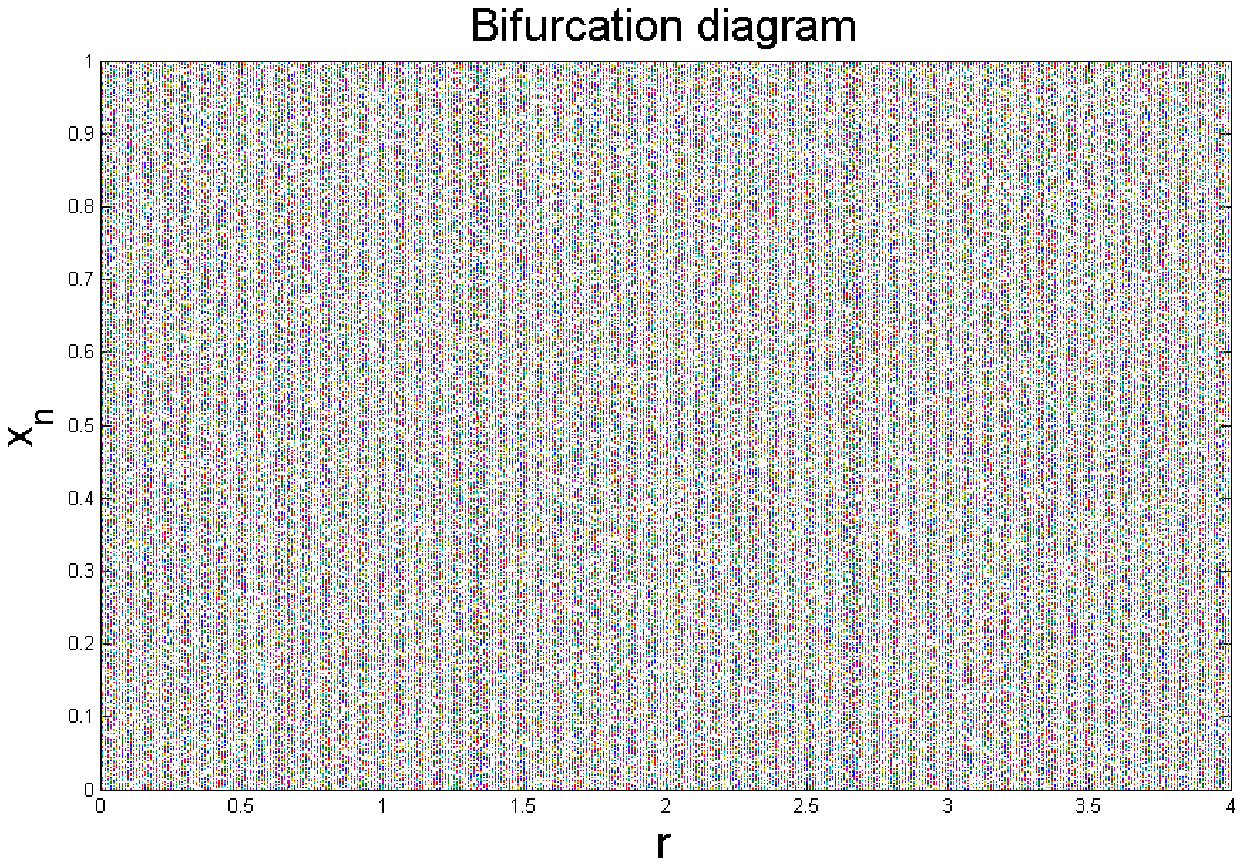}}
\hspace{2mm}
\subfigure[{}]{\label{f4-d}
\includegraphics*[width=.35\textwidth]{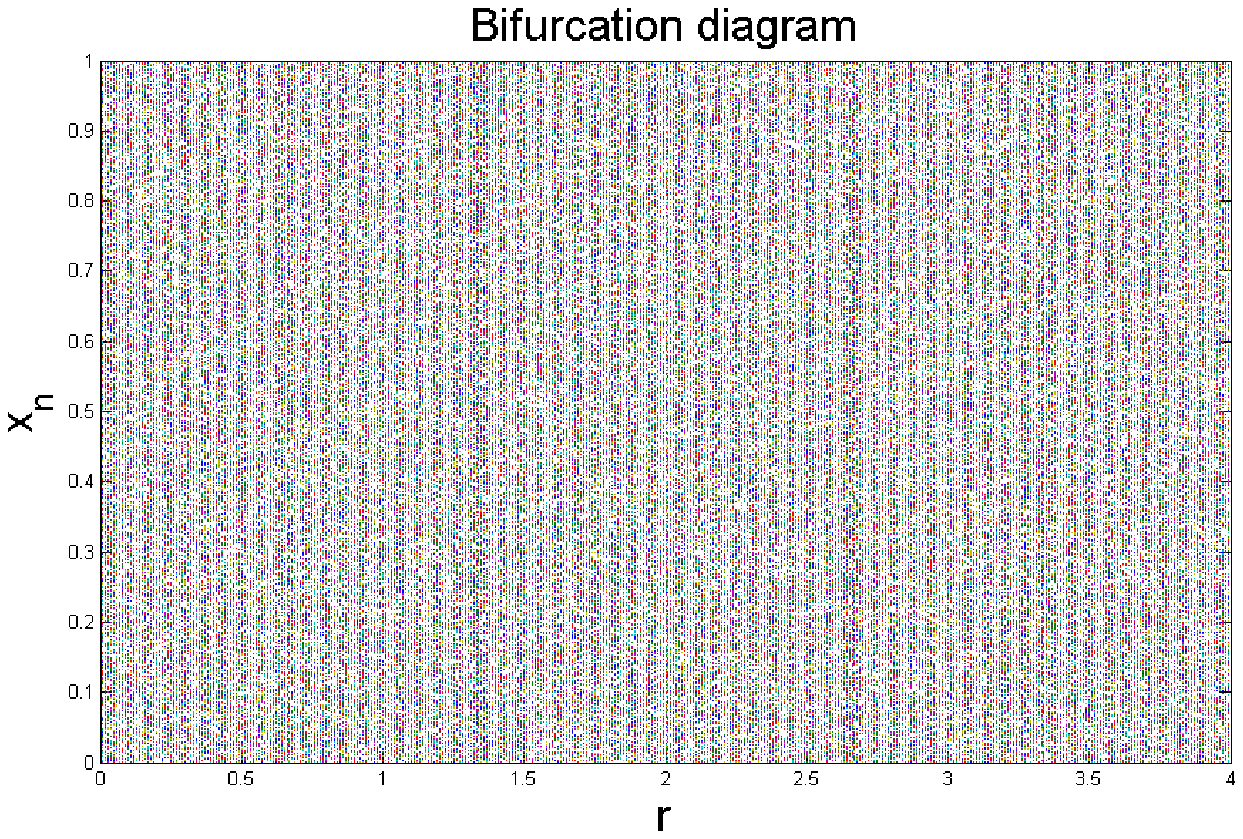}}
\caption{{\footnotesize
Bifurcation diagram for the (a) Logistic Tent map, (b) Case (i),
(c) Case (ii), (d) Case(iii).
}}
\end{figure}
\begin{figure}[!ht]
\centering
\subfigure[{}]{\label{f5-a}
\includegraphics*[width=.35\textwidth]{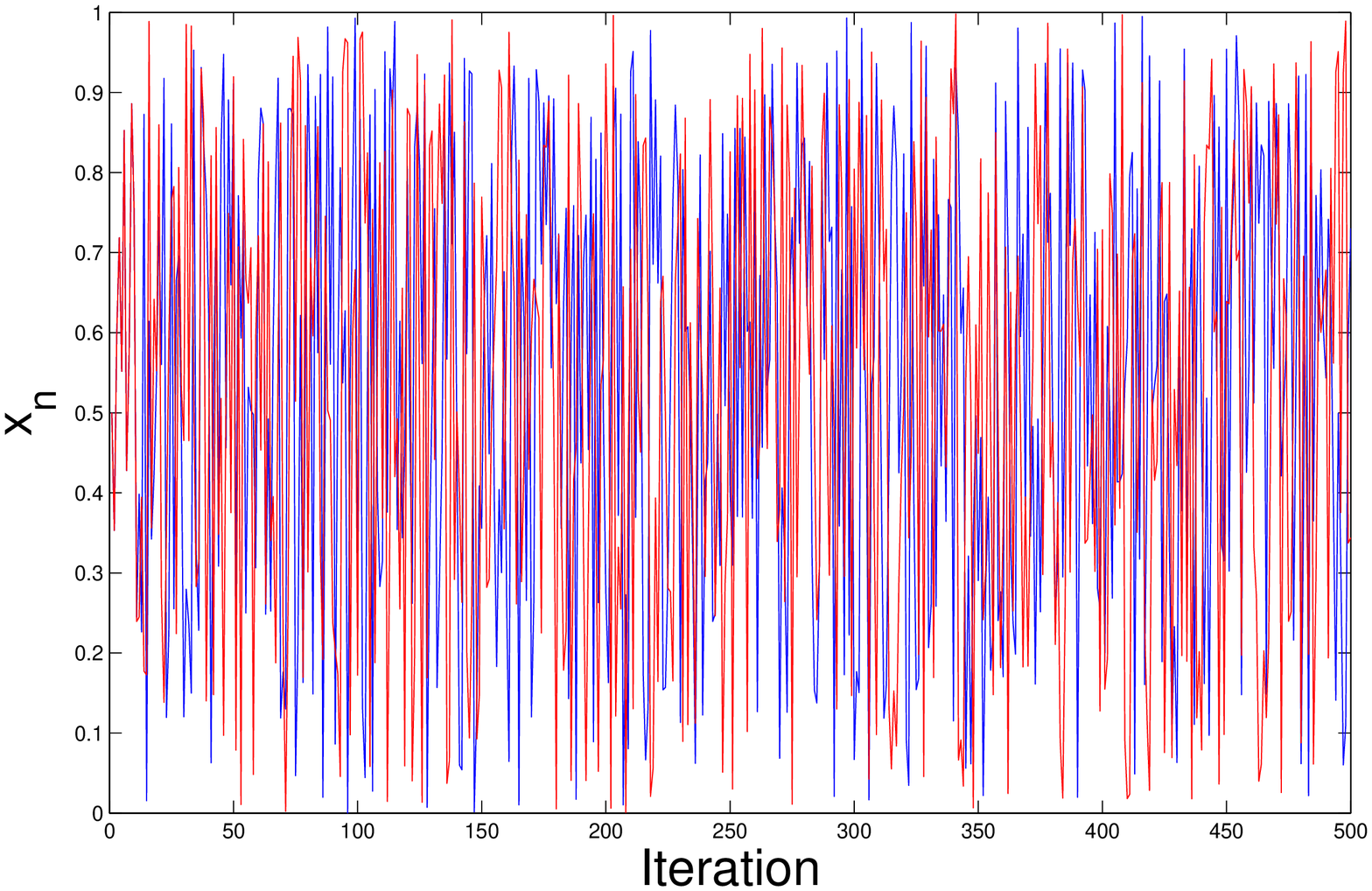}}
\hspace{2mm}
\subfigure[{}]{\label{f5-b}
\includegraphics*[width=.35\textwidth]{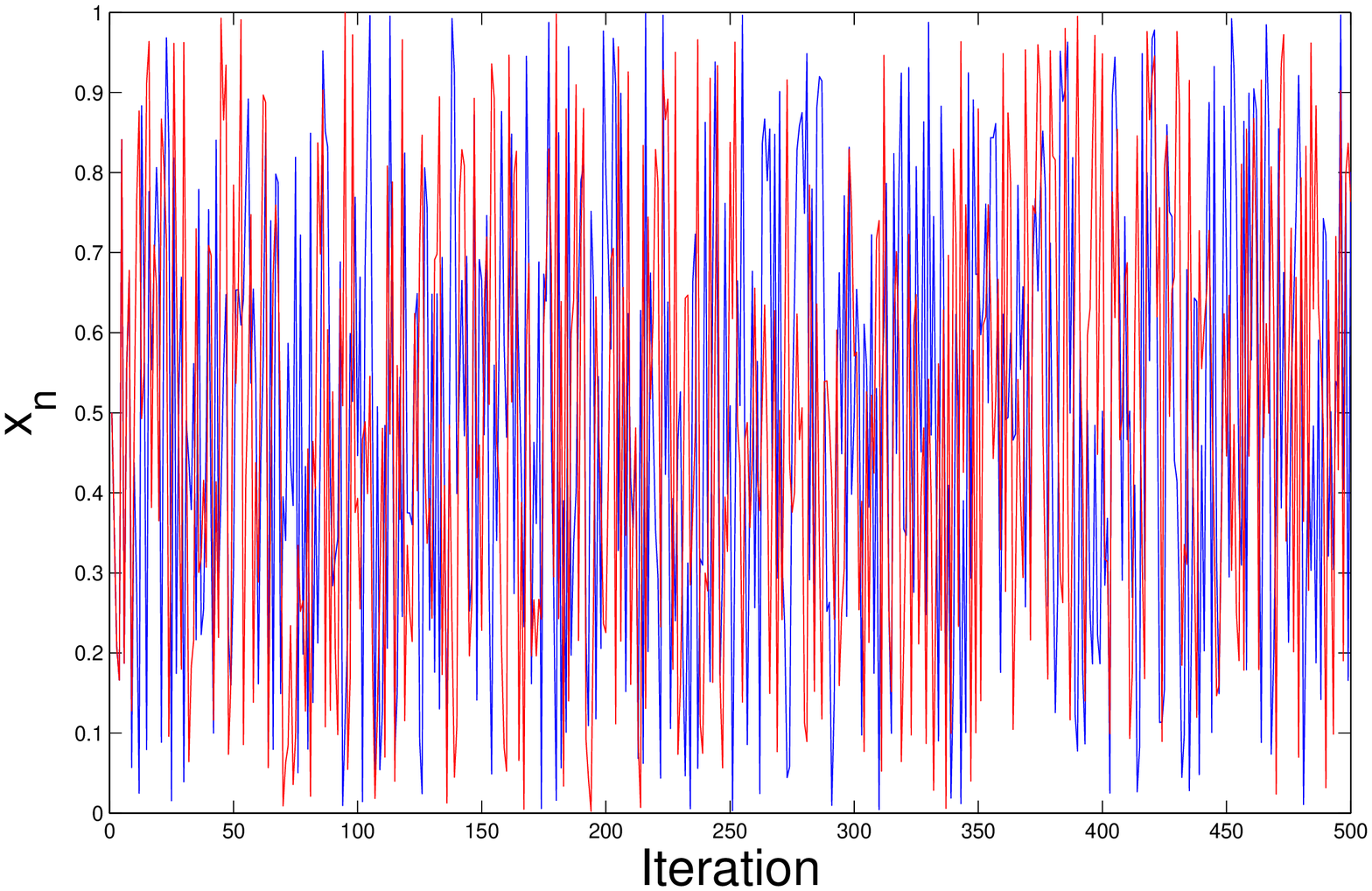}}
\caption{{\footnotesize (a) Case (i) with $x_0=0.5$ (blue) and $x_0=0.5+10^{-15}$ (red); (b) Case (ii) with $x_0=0.5$ (blue) and $x_0=0.5+10^{-15}$ (red).}
}
\label{f5}
\end{figure}
\section{Proposed encryption and decryption process}
In this section, we introduce some details about the proposed encryption and decryption algorithm.
\subsection{Encryption process}
We assume that the size of the input color image is $m \times n $. The encryption process is written in the following steps.\\
\begin{center}
\begin{figure}
\centering
\includegraphics[width=1.1\textwidth]{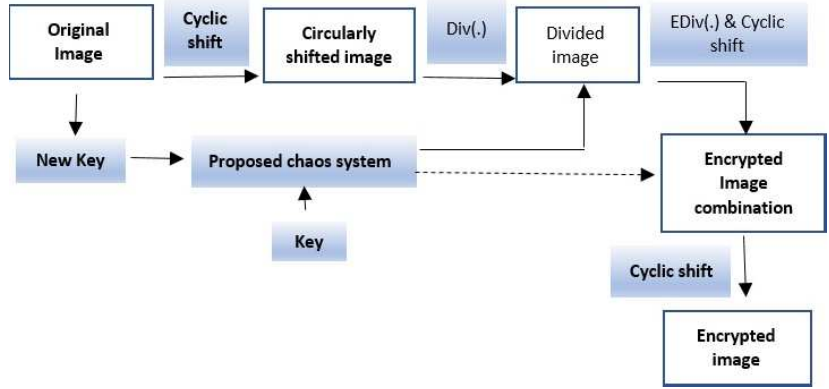}
\caption{{\footnotesize
The encryption process of the proposed algorithm.
} }
\label{f11}
\end{figure}
\end{center}

\textbf{Step~1. } For the color plain image A, we determine $y_k,y^0_k~(k=1,2,3)$ as follows
\begin{align*}
&y_k:=\sum_{i,j}A(i,j,k),~ k=1,2,3,\\
&y^0_k:=\frac{y_k}{m\times n\times 255},~ k=1,2,3.
\end{align*}
Therefore in this step we can find the different secret keys for different plain images.
Also to generate different keys in each iteration, we consider $y^0_0$ and $y^0_4$ as random
numbers in $[0,1]$. For gray scale image or binary image, we consider $y_1:=\sum_{i,j}A(i,j)$,
$y^0_1:=\frac{y_1}{m\times n\times 255}$ ,$y^0_0:=rand()$  and $y^0_2:=rand()$. Also
for  gray scale image or binary image $y_k, y^0_k~(k=3,4)$ are not used in the  encryption algorithm.\\

\textbf{Step~2. }  By using cyclic shift operation we  find
\begin{align*}
&A=circshift(A,[\lfloor \frac{y_1}{4} \rfloor~~-\lfloor\frac{y_2}{4}\rfloor~~\lfloor\frac{y_3}{4}\rfloor]),\\
&A=circshift(A,[\lfloor \frac{y_1+y_2}{10} \rfloor~~\lfloor\frac{y_2+y_3}{10} \rfloor~~
-\lfloor \frac{y_3+y_1}{10} \rfloor]),
\end{align*}
where $ circshift(A,r)$ circularly shifts the values in array $A$ by $r$ positions.
For gray scale image or binary image, this step has been changed as follows
\begin{align*}
&A=circshift(A,[\lfloor \frac{y_1}{4} \rfloor~~-\lfloor\frac{y_1}{3}\rfloor]),\\
&A=circshift(A,[\lfloor \frac{2y_1}{10} \rfloor~~\lfloor\frac{2y_1}{5}\rfloor]).
\end{align*}

\textbf{Step~3.}
In this step, by using $Div(A)$ image $A$ is divided
into twelve parts. In the $Div(A)$  function, image is divided into three color (red, green, blue)
then each part is divided into four equal parts. However in the encryption algorithm, other division functions can be used.
We can find
\begin{align*}
\{A^r_{i},A^g_{i},A^b_{i}\}_{i=1}^{4}=Div(A).
\end{align*}

\textbf{Step~4.}
We define $\Lambda(y^0,r,n)$
as follows
\begin{align}
\Lambda(y^0,r,n):=(x_0,x_1,\ldots,x_{n-1}),
\end{align}
where $x_i$ $(i=1,\ldots,n-1)$ are defined by using (\ref{e1}) with $x_0=y^0$ and parameter $r$.
Also we consider $\Omega(K_i,r)$ for all $i\in\{1,2,3,4\}$ as a matrix with the following elements
\begin{align}
&\Omega(K_i,r)(1,:):=\Lambda(y^0_i,r_i,\lfloor n/2\rfloor),~i=1,3,\\
&\Omega(K_i,r)(1,:):=\Lambda(y^0_i,r_i,n-\lfloor n/2\rfloor),~i=2,4,\\
&\Omega(K_1,r)(2:\lfloor m/2\rfloor,j):=\Lambda(\Omega(K_1,r_1)(1,j),r_1,\lfloor n/2\rfloor)^T,~j=1,\ldots\lfloor n/2\rfloor,\\
&\Omega(K_2,r)(2:\lfloor m/2\rfloor,j):=\Lambda(\Omega(K_2,r_2)(1,j),r_2,n-\lfloor n/2\rfloor)^T,~j=1,\ldots,n-\lfloor n/2\rfloor,\\
&\Omega(K_3,r)(2:m-\lfloor m/2\rfloor,j):=\Lambda(\Omega(K_3,r_3)(1,j),r_3,\lfloor n/2\rfloor)^T,~j=1,\ldots\lfloor n/2\rfloor,\\
&\Omega(K_4,r)(2:m-\lfloor m/2\rfloor,j):=\Lambda(\Omega(K_4,r_4)(1,j),r_4,n-\lfloor n/2\rfloor)^T,~j=1,\ldots,n-\lfloor n/2\rfloor.
\end{align}
(a)~ For $i=1,\ldots,4$ we find
\begin{align*}
&K_i:=\Lambda(y^0_i,r_i,\lfloor \frac{n}{2} \rfloor),\\
&t_i:=mod(K_i(1,\lfloor \frac{n}{2} \rfloor)\times10^5,m\times n).
\end{align*}
Also for $i=0$ we define
\begin{align*}
&t_0:=\lfloor \frac{\sum^{4}_{i=1}t_i}{4}\rfloor,\\
&K_0:=\Lambda(y^0_0,r_0,m\times n).\\
\end{align*}

(b)~ We define $Z^1_i, Z^2_i~(i=0,\ldots,4$) as
\begin{align*}
&Z^1_i:=\Omega(K_i,r^i)\times 255,\\
&Z^2_i:=circshift(Z^1_i,t_i).
\end{align*}
Also $Z^j_0~(j=1,2)$ are defined as follows
\begin{flushleft}
$for~i=1:m~do$\\
$~~~~for~ j=1:n~do$\\
$~~~~~~Z^1_0(i,j):=K_0(1,j+(i-1)n)\times255,$\\
$~~~~end$\\
$end$\\
\end{flushleft}
\begin{align*}
Z^2_0:=circshift(Z^1_0,-t_0),
\end{align*}

(c)~By using bitxor operation
matrixs $A^j_i (i=1,2,3,4,~j=r,g,b)$ are encrypted
as  follows

\begin{flushleft}
$for~i=1:4 do$\\
$~~~~for~ j=r, g, b~do$\\
$~~~~~~~~for~w=1:2 do$\\
$~~~~~~~~~~DA^j_i:=bitxor(A^j_i,Z^w_i),$\\
$~~~~~~~~end$\\
$~~~~end$\\
$end$\\
\end{flushleft}

\textbf{Step~6.}
In the  encryption algorithm, the inverse function of the $Div(\cdot)$ is shown by $EDiv(\cdot)$.\\

(a)~By using the inverse function of the $Div(\cdot)$ we find
\begin{align*}
EA:=EDiv(\{DA^r_{i},DA^g_{i},DA^b_{i}\}_{i=1}^{4}),
\end{align*}
where $EA$ denotes the joined encrypted image. \\

(b)~We transform matrix $EA$ as follows
\begin{align*}
&EA(:,:,1)=circshift(EA(:,:,1),[-y_1+t_0 ~~t_0]),\\
&EA(:,:,2)=circshift(EA(:,:,2),[t_0~~ -y_2+t_0]),\\
&EA(:,:,3)=circshift(EA(:,:,3),[t_0 ~~-y_3+t_0]).
\end{align*}

(c)~The final encrypted image is found as follows
\begin{flushleft}
$for~i=1:3$\\
$~~~~~~EA(:,:,i)=bitxor(EA(:,:,i),Z^1_0),$\\
$~~~~~~EA(:,:,i)=bitxor(EA(:,:,i),Z^2_0),$\\
$end$
\end{flushleft}

\begin{align}\label{enc1}
&EA=circshift(EA,[y_1+y_2~~y_2+y_3~~y_1+y_3]),\\\label{enc2}
&EA=circshift(EA,[\lfloor \frac{y_1}{50} \rfloor~~\lfloor\frac{y_2}{50}\rfloor~~-\lfloor\frac{y_3}{50}\rfloor]).
\end{align}
For gray scale image or binary image, Eq.s (\ref{enc1})-(\ref{enc2})  has been changed as follows

\begin{align*}
&EA=circshift(EA,[y_1~~-2y_1),\\
&EA=circshift(EA,[\lfloor \frac{y_1}{50} \rfloor~~-\lfloor\frac{y_1}{35}\rfloor]).
\end{align*}
\subsection{Decryption process}
The decryption process is the inverse process of the encryption process,
so we remove the decryption details.
\section{Simulation results and security analysis}
\subsection{Simulation results}
In order to illustrate the encryption algorithm results and security analysis, we consider some test
problems.
In this section in all figures we use Case (ii) in the encryption process.
We have computed the numerical results by MATLAB 7.11.0
programming.
As plain images, we use color images ``Sailboat on lake" ,  ``Airplane (F-16)"  ($512\times 512\times 3$ pixels),
 ``Lena" ($256\times 256\times 3$, $512\times 512\times 3$  pixels) and gray scale image ``Lena" ($256\times 256$ , $512\times 512$  pixels), ``girl" ($256\times 256$ pixels).
 Also	``Horse"  ($444 \times 455$ pixels) is used as binary image.
The simulations results are given in Figs. \ref{f5}-\ref{f9}.
In Figs. \ref{f5}-\ref{f6} we show  the encryption algorithm results.
The binary image has  only two pixel values 0 and 1, therefore
it is  a difficult case for encryption.
As can be seen in
the Fig.\ref{f6},
by using the proposed algorithm  for
the binary image, the image is changed to  a noise-like encrypted image with
flat histogram.
The running time of encryption and decryption are shown in Table \ref{table:trrr}.
\begin{center}
\begin{table}[ht!]
\caption{ {\footnotesize Encryption and decryption times of the color images by using Case (ii).}}
\label{table:trrr}
\centering
\begin{tabular}{cccccccccccc}
\hline
Image  & Encryption time &Decryption time \\
\hline
Lena  (512 $\times$ 512)      &1.130 s & 1.044 s     \\
Sailboat on lake                  &1.105 s  &1.044 s\\
Airplane (F-16)                 &1.139 s  &1.024 s  \\
\hline
\end{tabular}
\end{table}
\end{center}
\begin{center}
\begin{figure}
\centering
\includegraphics[width=1.1\textwidth]{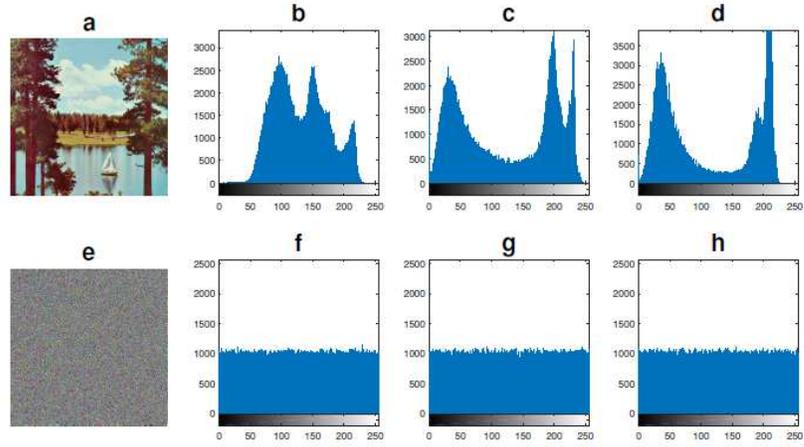}
\caption{{\footnotesize (a) Original image, (b-c-d)  Histograms of the R, G, B components of the original image, (e) Encrypted image, (f-g-h) Histograms of the R, G, B components of the encrypted image (Sailboat on lake).}}
\label{f5}
\end{figure}
\end{center}
\begin{center}
\begin{figure}
\centering
\includegraphics[width=1.1\textwidth]{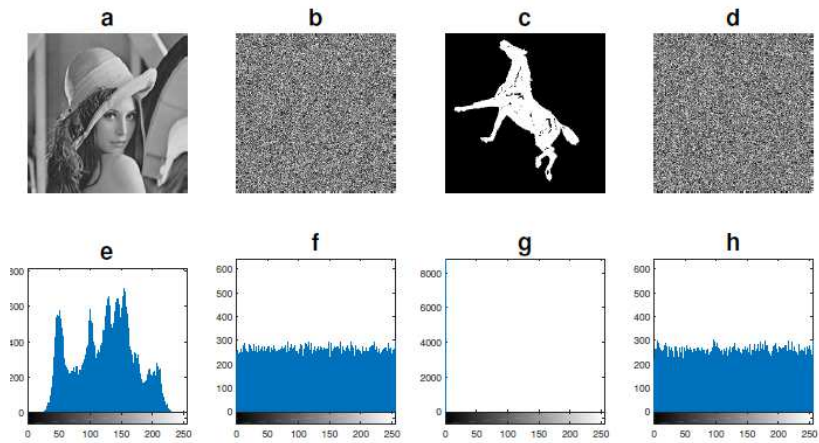}
\caption{{\footnotesize (a-b) Original and encrypted image, (e-f) Histograms of  the original image  and encrypted image (Lena -$256\times 256$), respectively, (c-d) Original and encrypted image, (g-h) Histograms of  the original image  and encrypted image (Horse), respectively.}}
\label{f6}
\end{figure}
\end{center}

\subsection{Security analysis}
In this subsection, the different security parameters are discussed, for details see \cite{11,12}.
\subsubsection{Security key analysis}
\begin{center}
\begin{figure}[!ht]
\centering
\includegraphics[width=1.1\textwidth]{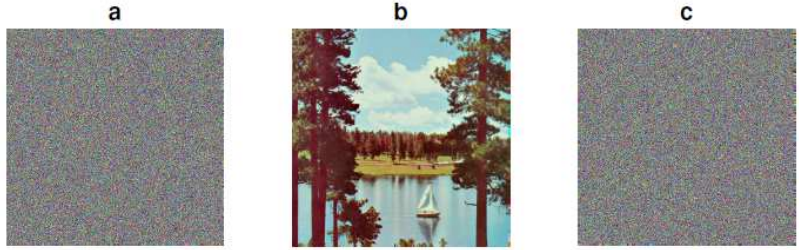}
\caption{{\footnotesize (a) Encrypted image (Sailboat on lake), (b) Decrypted image by using $K$,  (c) Decrypted image by using $K_1$.}}
\label{f7}
\end{figure}
\end{center}

From the proposed algorithm, we can say that
the security keys of the  algorithm are composed of
ten parameters $\{r_i\}^{4}_{n=0}$  and $\{y^0_i\}^{4}_{n=0}$,
The range for $r_i ~(i=0,\ldots4)$ are $(0,4]$
Also $y^0_i ~(i=0,\ldots4)$ are in range of $[0,1]$.
If in the image encryption algorithm we consider the precision as $10^{-15}$
the key space is almost $10^{150}$, and this space
is sufficiently large to resist
the brute force attack.\\
To show key sensitivity, in the encryption process
we consider key as follows (this key is used to obtain all results)
\begin{align}
k:=(r_0,r_1,r_2,r_3,r_4)=(2,1,2,3.5,1.75).
\end{align}
In the decryption process, we use a small change for $r_0$ as
$r_0=2+10^{-15}$, we consider new key as $K_1$.
Simulation results in Fig. \ref{f7} show that by using new key we can not reconstruct the original image.
Therefore, the proposed algorithm has high key sensitivity.

\subsubsection{Statistical analysis}
\begin{center}
\begin{figure}
\centering
\includegraphics[width=1.1\textwidth]{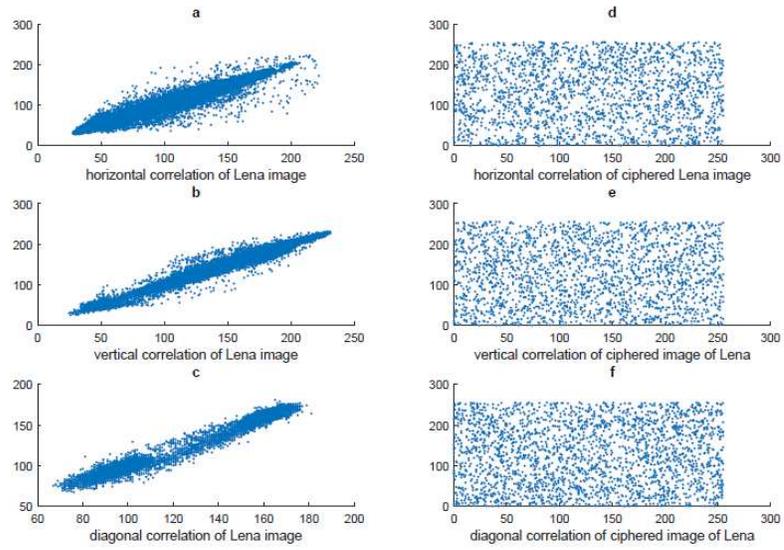}
\caption{{ \footnotesize
Correlation of neighborhood pixels at different directions before and after encryption 
of Lena ( 512 $\times$ 512 pixels).
} }
\label{fC}
\end{figure}
\end{center}

In this section, we study statistical analysis as part of the security analysis.
It is clear that a good encrypted image should be unrecognized hence the correlation values of
a good encrypted image are close to zero. Table \ref{table:t1} shows the correlation values for the original images
and encrypted images. We have found the correlation values using formula
 \cite{15}
\begin{align}
C_{xy}=\frac{E[(x-\mu_x)(y-\mu_y)]}{\sigma_x \sigma_y},
\end{align}
where $E[\cdot]$ denotes the expectation value, $\mu$ is
the mean value and $\sigma$ represents standard deviation.
From Table \ref{table:t1}, we can see that the original images have high
correlation values  while encrypted
images have very low correlation values. Also  correlation distributions for the original image and encrypted image of Lena image are shown in Fig.\ref{fC}. \\
\begin{center}
\begin{table}[ht!]
\caption{{\footnotesize Correlation coefficients of the encrypted color image of ''Sailboat on lake''.}}
\label{table:t1}
\centering
\begin{tabular}{cccccc}
\hline
Chaos map  &Component         &Horizontal      &Vertical &Diagonal        &Diagonal \\
              &                         &                    & &\tiny{ (lower left to top right)}&\tiny{(lower right to top left)}\\
\hline \hline
                      &R     &-0.0027      &-0.0003   &-0.0016       &0.0019     \\
 Case (i)             &G     &0.0009      &-0.0019    &-0.0008      &0.0027    \\
                      &B    &-0.0017      &-0.0005    &-0.0019      &0.0024   \\
\hline
                      &R     &0.0007      &-0.0003    &-0.0006       &-0.0002     \\
Case (ii)             &G     &0.0029      &-0.0005    &-0.0003       &-0.0003    \\
                      &B    & -0.0007      &0.0009    &0.0010      &-0.0023    \\
\hline
                       &R     &0.0003      &0.0018    &0.0020       &0.0027     \\
Case (iii)             &G     &0.0029      &-0.0005    &-0.0028       &0.0030    \\
                       &B    &-0.0007      &-0.0021    &0.0007      &-0.0021    \\

\hline
                        &R    &-0.0029      &0.0011   &0.0028         &-0.0005    \\
Logistic Tent           &G     &-0.0009     &0.0033    &0.0035         &-0.0002     \\
                        &B  &-0.0009     &0.0025    &0.0012         &0.0005\\
\hline
                        &R   &0.9558      & 0.9541    &0.9373       &0.9420     \\

Original image          &G    & 0.9715      &0.9663    &0.9510          &0.9530    \\
                        &B    &0.9710     & 0.9694    &0.9512        &0.9530\\
\hline
\end{tabular}
\end{table}
\end{center}
\begin{center}
\begin{table}[ht!]
\caption{{\footnotesize 
Comparison of Correlation coefficients of  image of  ``Lena'' ($512\times 512$).}}
\label{table:t1}
\centering
\begin{tabular}{cccccc}
\hline
Method    &Horizontal       &Vertical    &Diagonal    \\
           
\hline \hline
                    
 The proposed            &-0.0036     &-0.0020      &-0.0026     \\
                     
\hline
                    
Method in \cite{14}        &0.0139         &0.0073    &0.0104          \\

\hline
\end{tabular}
\end{table}
\end{center}

The other test is the information entropy.
The values of entropy are in range of $[0,8]$.
This test is used for evaluating the randomness of
an image. If the value of entropy of encrypted image is
close to the maximum value means the excellent random property.
The information entropy is defined as follows \cite{12}
\begin{align}\label{ex1}
H(k)=-\sum^{w-1}_{i=0}P(k_i)\log_2P(k_i),
\end{align}
where $w$ is the gray level and $P(\cdot)$ denotes the probability of symbol.
Results for the information entropy are tabulated in Table \ref{table:t2}.
From Table \ref{table:t2}, we can say that test results for the encryption algorithm
are close to the maximum value.
Also Figs. \ref{f5}-\ref{f6} show that the histograms of plain images are not flat
while the histograms of encrypted  images are in flat distributions.\\

From the above discussion, it can be concluded that the proposed algorithm
has stronger ability to resist statistical attacks.
\begin{center}
\begin{table}[ht!]
\caption{{\footnotesize Informaion entropies of the encrypted color image.}}
\label{table:t2}
\centering
\begin{tabular}{cccccccccccc}
\hline
Chaos map              &Image            & &R      & &G    & &B       \\
\hline \hline
                            &Lena  ($512\times 512\times 3$)                         &&7.9970       & & 7.9972    &&7.9970           \\
Case (i)                &Sailboat on lake          &&7.9993        && 7.9993      && 7.9992             \\
                            &Airplane (F-16)           &&7.9994      && 7.9992 && 7.9993    \\
\hline
                                &Lena ($512\times 512\times 3$)                     &&7.9967                   & & 7.9967     &&7.9972            \\
Case (ii)                 &Sailboat on lake         &&7.9992        && 7.9993      && 7.9993             \\
                             &Airplane (F-16)     &&7.9994                    && 7.9993       && 7.9994    \\
\hline
                            &Lena ($512\times 512\times 3$)                         &&7.9970       & & 7.9973     &&7.9975            \\
Case (iii)                &Sailboat on lake          &&7.9994        && 7.9993      && 7.9993             \\
                            &Airplane (F-16)           &&7.9991       && 7.9992 && 7.9994    \\
\hline
                              &Lena  ($512\times 512\times 3$)                           &&7.9975       & & 7.9970     &&7.9970            \\
Logistic Tent            &Sailboat on lake          &&7.9993        && 7.9992      && 7.9993             \\
                              &Airplane (F-16)           &&7.9993       && 7.9994 && 7.9994    \\
\hline
\end{tabular}
\end{table}
\end{center}
\subsubsection{Sensitivity analysis}
NPCR (Number of Pixels change Rate) denotes the number of pixels change rate while one pixel of plain image changed.
Also UACI (Unified Average Changing Intensity) measures the average intensity of differences
between the plain image and encrypted image.
The ideal value for NPCR is $100\%$ while the ideal value for UACI is $33.\overline{33}\%$.
When the value of NPCR gets closer to $100\%$,
 the encryption algorithm is
more sensitive to the changing of
plain image, therefore the algorithm can effectively resist plaintext attack.
Also when the value of NPCR gets closer to $33.\overline{33}\%$,
the proposed algorithm can effectively resist differential attack.
In computing, we consider NPCR and UACI as follows
\begin{align}\label{ex1}
&NPCR=\frac{\sum_{i,j}D(i,j)}{m\times n}\times 100\%,\\
&UACI=\frac{1}{m\times n}\big [ \frac{\sum_{i,J}|C_1(i,j)-C_2(i,j)|}{255}\big]\times 100\%,
\end{align}
where
\begin{align}\label{ex1}
&D(i,j):=\left\{%
\begin{array}{ll}
1,&when~C_1(i,j)\neq C_2(i,j),\\
\\
0,&when~C_1(i,j)= C_2(i,j).\\
\end{array}%
\right.
\end{align}
In above formulae, $C_1(i,j)$ and $C_2(i,j)$ are denoted  the cipher
image before and after one pixel of the plain image
is changed. In the Lena, Sailboat  on lake and Airplane images,  $A(200,150,1)$, $A(100,250,2)$ and $A(180,334,3)$ are changed to 0,
respectively. As can be seen in Table \ref{table:t3} , results are close to ideal values.
\begin{center}
\begin{table}[ht!]
\caption{ {\footnotesize UACI and NPCR of the encrypted color image.}}
\label{table:t3}
\centering
\begin{tabular}{cccccccccccc}
\hline
Image  &\multicolumn{3}{c}{UACI }&\multicolumn{3}{c}{NPCR }      \\
           &R       &G     &B                  &R      &G     &B     \\
\hline \hline
Lena  ($512\times 512\times 3$)     &33.4457   & 33.5589     &33.5243       &99.6078       &99.6140   &99.6033     \\
Sailboat on lake                &33.4617   &33.3928      &33.5061             &99.6082      &99.6231   &99.6048   \\
Airplane (F-16)                 &33.4744   &33.4482      & 33.4813
   &99.6353        &99.6059  &99.6021     \\
\hline
\end{tabular}
\end{table}
\end{center}
\subsubsection{Noise and Data loss  attacks }
In continuation of discussion, by using the simulation results we study noise and data loss attacks.
In real applications, a part of the encrypted image may be lost during transmission.
A proper encryption algorithm
should resist the data loss and noise attacks.
In Fig. \ref{f10}, we remove $200 \times 200$ of encrypted image. In fact we consider $A(80:280,80:280)=0_{200\times 200}$. Also in Fig. \ref{f11}, for encrypted image we add Gaussian noise with zero-mean and different
variance. After the decryption process,
we can see that the reconstructed images contain most of original visual in formation
and we can recognize the original image from decrypted image.
\begin{center}
\begin{figure}
\centering
\includegraphics[width=1.1\textwidth]{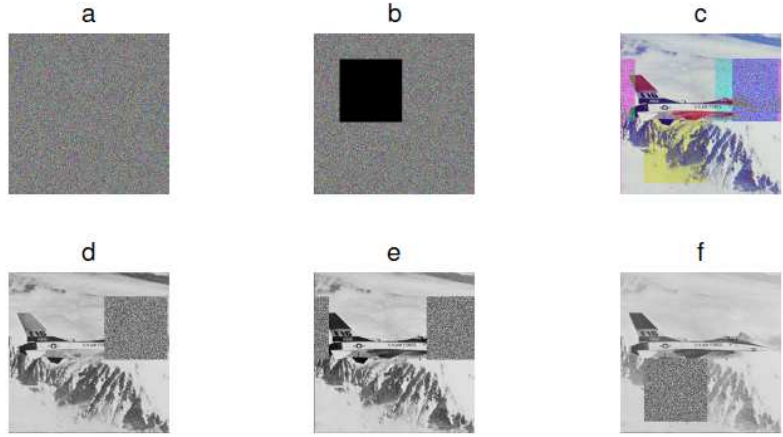}
\caption{{\footnotesize
Cropped attack experiment: (a) Encrypted image, (b) cropped attack image,
(c) Decryption result, (d-e-f) Decryption result for the R, G, B components, respectively.
} }
\label{f10}
\end{figure}
\end{center}
\begin{center}
\begin{figure}
\centering
\includegraphics[width=1.1\textwidth]{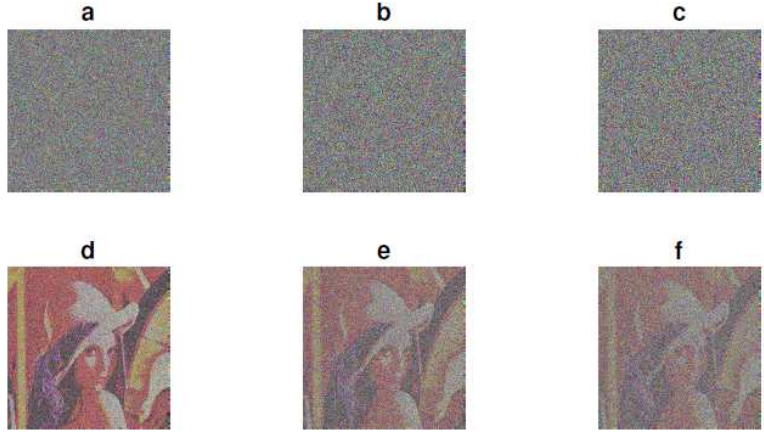}
\caption{{ \footnotesize
Noise attack experiment with Gaussian noise (a-b-c) noise attack image with var=0.1,0.4,0.8, respectively, (d-e-f) decryption
with var=0.1,0.4,0.8, respectively.
} }
\label{f11}
\end{figure}
\end{center}
\subsubsection{Chosen-plain text attack}
\begin{center}
\begin{figure}
\centering
\includegraphics[width=1.1\textwidth]{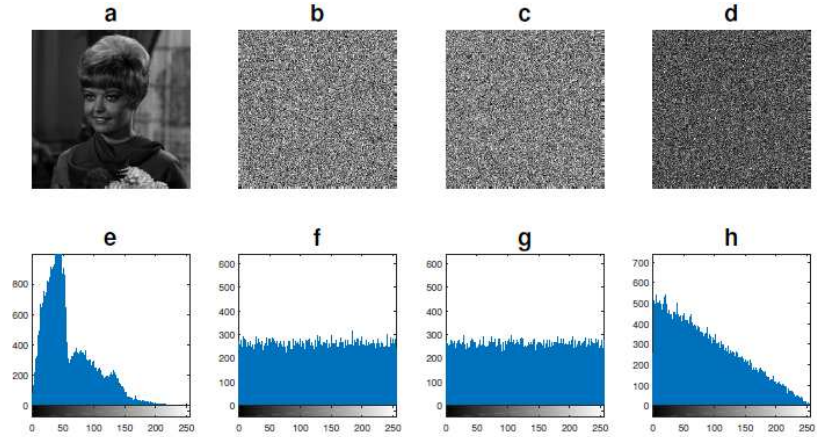}
\caption{{\footnotesize
(a-e) The original image and its histogram, (b-f) the first encrypted image and its histogram, (c-g) the second encrypted image and its histogram,
(d-h) the pixel-to-pixel difference and its histogram.
}}
\label{f9}
\end{figure}
\end{center}

In  the  encryption algorithm, in Step 1, random numbers are used.
So this algorithm create different
encrypted image each time when the encryption algorithm is applied
to the same image.
In Fig. \ref{f9} we use $K$ as key and we run the encryption algorithm twice.
The first and second encrypted image are consider as  $C_1$ (Fig. \ref{f9}(b)) and $C_2$ (Fig. \ref{f9}(c))
, respectively. To illustrate the difference between the two images, we use
pixel-to-pixel difference as $|C_1-C_2|$.
As can be seen in Fig. 10, two encrypted images are different.
Then  the our algorithm is able to withstand the chosen-plain text attack.

\section{Conclusion}
In this paper, we have constructed a new combination chaotic system based on Logistic and Tent systems.
By using this  new combination chaotic system  a large number of chaotic map can be produced. Also we have proposed
a new image encryption algorithm based on combination chaotic system. It is shown that
the proposed encryption algorithm can effectively resist differential, statistical, noise, data loss, chosen-plain text attacks.


\begin{thebibliography}{99}
\providecommand{\doi}[1]{DOI~\discretionary{}{}{}#1}
\bibitem{1}
R.~Matthews, {\em On the derivation of a chaotic encryption algorithm},
Cryptologia 4 (1989) 29-42.
\bibitem{2}
{\"U}. {\c C}avu{\c s}o{\u g}lu, S. Ka{\c c}arb, I. Pehlivanb, A. Zengina, {\em Secure image encryption algorithm design using a novel chaos based S-Box},
Chaos, Solitons $\&$ Fractals 95 (2017) 92-101.
\bibitem{3}
A. Kanso, M. Ghebleh, {\em An algorithm for encryption of secret images into meaningful images},
Optics and Lasers in Engineering 90 (2017) 196-208.
\bibitem{4}
X.Y.  Wang,  L. Yang, R.Liu, A. Kadir, {\em A chaotic image encryption algorithm based on perceptron
model}, Nonlinear Dynamics  62 (2010) 615-621
\bibitem{5}
G. Gu, J. Ling,
{\em A fast image encryption method by using chaotic 3D cat maps}, Optik - International Journal for Light and Electron Optics 125 (2014) 4700-4705.
\bibitem{6}
M. Kumar, A. Vaish, {\em Encryption of color images using MSVD in DCST domain}, Optics and Lasers in Engineering 3 (1990) 278-285.
\bibitem{7}
M. Ausloos, M. Dirickx, {\em The Logistic Map and the Route to Chaos: From the Beginnings to Modern Applications}, Springer 2006.
\bibitem{8}
R. Hilborn,
{\em Chaos and Nonlinear Dynamics: An Introduction for Scientists and Engineers}, Oxford University Press 2000.
\bibitem{9}
S.C. Satapathy, A. Govardhan, K.S. Raju, J.K. Mandal, {\em Emerging
ICT for Bridging the Future - Proceedings of the  49th Annual Convention of the Computer Society of India (CSI)},
Springer International Publishing Switzerland 2015.
\bibitem{10}
S. Lynch,  {\em Dynamical Systems with Applications using MATLAB $\circledR$}, Second Edition, Birkh{\"a}user Boston 2014.
\bibitem{11}
A. Uhl, A.  Pommer, {\em Image and Video Encryption From Digital Rights Management to Secured Personal Communication}, Springer	2004.
\bibitem{12}
F.A El-Samie, H.H. Ahmed, I.F. Elashry, M.H. Shahieen, O.S. Faragallah, E.M. El-Rabaie, S.A. Alshebeili,
{\em Image Encryption: A Communication Perspective}, CRC Press 2014.
\bibitem{15}
N. Zheng, J. Xue, {\em Statistical learning and pattern analysis for image and video processing}, Springer Science $\&$ Business Media, 2009.
\bibitem{14}
G. Gu, J. Ling, {\em A fast image encryption method by using chaotic 3D cat maps}, Optik-International Journal for Light and Electron Optics, 17 (2014) 4700-4705.

\end{thebibliography}
\end{document}